\PassOptionsToPackage{svgnames}{xcolor}
\documentclass[11pt,reqno]{article}
\usepackage{amsmath,hyperref,amssymb, amsthm}
\usepackage{authblk}
\usepackage[all]{xy}
\usepackage{graphicx,url}
\usepackage{color}
\usepackage{slashed}
\usepackage{tikz}
\usepackage{lipsum}
\usepackage{caption}
\usepackage{enumitem}
\usepackage[style=ieee]{biblatex}
\newcommand{\be}{\begin{eqnarray}}
\newcommand{\ee}{\end{eqnarray}}
\newcommand{\eeq}{\end{equation}}
\newcommand{\beq}{\begin{equation}}

\allowdisplaybreaks \numberwithin{equation}{section}

\DeclareSymbolFont{AMSa}{U}{msa}{m}{n}
\DeclareSymbolFont{AMSb}{U}{msb}{m}{n}
\DeclareMathSymbol{\fieldR}{\mathalpha}{AMSb}{"52}

\newcommand{\CA}{{\cal A}}

\newcommand{\CH}{\mathcal{H}}
\newcommand{\Hh}{\mathcal{H}}

\newcommand{\CI}{{\cal I}}
\newcommand{\CL}{{\cal L}}
\newcommand{\CN}{\mathcal{N}}
\newcommand{\CM}{\mathcal{M}}

\newcommand{\calC}{\mathcal{C}}
\newcommand{\calR}{\mathcal{R}}
\newcommand{\CT}{\mathcal{T}}

\newcommand{\dual}{\star}
\newcommand{\nL}{\mathsf{L}}

\newcommand{\HRR}{\text{RR}}

\newcommand{\GTVW}{\text{GTVW}}

\newcommand{\NN}{\mathbb{N}}
\newcommand{\ZZ}{\mathbb{Z}}
\newcommand{\RR}{\mathbb{R}}

\newcommand{\QQ}{\mathbb{Q}}

\newcommand{\cTop}{\mathsf{Top}}

\def\beq{\begin{equation}}
\def\eeq{\end{equation}}
\def\bea{\begin{eqnarray}}
\def\eea{\end{eqnarray}}

\def\<{\langle}

\newtheorem{theorem}{Theorem}
\newtheorem{claim}[theorem]{Claim}
\newtheorem{conjecture}[theorem]{Conjecture}

\usepackage{tcolorbox}
\tcbuselibrary{skins,breakable}
\usetikzlibrary{shadings,shadows}
    {\endtcolorbox}

    {\endtcolorbox}

    {\endtcolorbox}

\title{Towards a classification of topological defects in $K3$ sigma models}
\author[1]{Roberta Angius\thanks{roberta.angius@csic.es}}
\author[2,3]{Stefano Giaccari\thanks{s.giaccari@inrim.it}}

{\small \affil[1]{\small  Instituto de F\'{\i}sica Te\'orica
IFT-UAM/CSIC,  C/ Nicol\'as Cabrera 13-15, Campus de Cantoblanco, 28049 Madrid, Spain}
\affil[2]{\small Dipartimento di Fisica e Astronomia `Galileo Galilei', Universit\`a di Padova \& INFN, sez. di Padova, Via Marzolo 8, 35131, Padova, Italy}
\affil[3]{\small Istituto Nazionale di Ricerca Metrologica, Strada delle Cacce 91, I-10135 Torino, Italy}}
\date{July 2025}

\begin{document}

\maketitle
\abstract{Given a $K3$ surface, a supersymmetric non-linear K3 sigma
model is the internal superconformal field theory (SCFT) in a six dimensional compactification of
type IIA superstring on $\mathbb{R}^{1,5} \times K3$. These models have attracted attention due to the discovery of  
Mathieu moonshine phenomena for the elliptic genera of K3 surfaces, and have played a pivotal role in extending Mukai's theorem on classification of symplectic automorphisms of $K3$ surfaces.
We report on recent progress \cite{Angius:2024evd} in characterizing topological defects in $K3$ models, generalizing the notion of symmetries to categories of topological operators supported on arbitrary codimension submanifolds with
possibly non-invertible fusion rules. Taking advantage of the interpretation of Mukai lattice as the D-brane charge lattice, we present a number of general results for the category
of topological defect lines preserving the superconformal algebra and spectral flow, obtained by studying their fusion with boundary states.
While for certain K3 models infinitely many simple defects, and even a continuum,
can occur, at generic points in the moduli space the category is actually trivial, i.e. it is generated
by the identity defect. Furthermore,
if a K3 model is at the attractor point for some
BPS configuration of D-branes, then all topological defects have integral quantum dimension.
We also introduce a conjecture that a continuum of topological defects arises if and only if the K3 model is
a (possibly generalized) orbifold of a torus model. These general results are confirmed by the
analysis of significant examples. We also point out the connection to recent studies of topological defects in Conway moonshine module \cite{Volpato:2024goy,Angius:2025xxx,Angius:2025zlm}.

\section{Introduction}
The finite symmetry groups, i.e. the symplectic automorphisms,  of $K3$ surfaces have gained considerable attention ever since the seminal work of Nikulin \cite{nikulin1976finite,nikulin1980finite,nikulin1980integral}, following which  Mukai showed that the finite group of symplectic automorphisms of a $K3$ surface is isomorphic to a subgroup of the Mathieu group $\mathbb{M}_{23}$ with  at least five orbits on its defining action on $24$ elements, and in this way he was able to list eleven maximal subgroups \cite{Mukai:1988}.  This result was nicely proven by Kondo using a lattice-theoretic approach \cite{Kondo:1988}. In particular, he considered the unique unimodular integral lattice $\Gamma^{4,20}\subset \mathbb{R}^{4,20}$ of signature $(4,20)$, which is isomorphic to the even integral homology of $K3$. Then, denoting by $L^G$ the sublattice of $G$-invariant vectors of  $L\equiv \Gamma^{4,20}$ and $L_G$ its orthogonal complement, he characterized $G$ by considering the embedding $L_G(-1)$ into the $23$ Niemeier lattices with roots. On the physics' side, a renewed interest in the symmetries of $K3$ came from Eguchi, Ooguri and
Tachikawa's discovery of Mathieu moonshine \cite{Eguchi:2010ej}, i.e. the realization that the multiplicities with which the $\mathcal{N}=4$ characters appear in the elliptic genus of $K3$ are sums (with integer coefficients) of the dimensions of representations of the Mathieu group  $\mathbb{M}_{24}$. This evidence for a hidden $\mathbb{M}_{24}$ symmetry underlying the elliptic genus of $K3$ motivated the work of Gaberdiel, Hohenegger and Volpato \cite{Gaberdiel:2011fg} to generalize the lattice
approach of Kondo for K3 surfaces to non-linear sigma models (NLSMs) on $K3$. In this context, the integral homology lattice $\Gamma^{4,20}$  has a transparent physical interpretation as the lattice of D-brane charges, whose dual is embedded in the $24$ real dimensional space $V$ of R-R ground states with conformal weights $h=\tilde h=\frac{1}{4}$. A given $K3$ NLSM should then be thought of as associated with a choice of a positive definite $4$-dimensional plane $\Pi \subset V$. In physics's parlance $\Pi$ is the space spanned by the four spectral flow generators, i.e. the R-R ground states in a $(\frac{1}{4},\frac{1}{2};\frac{1}{4},\frac{1}{2})$ representation of $\CN=(4,4)$, whereas its orthogonal complement $\Pi^\perp\subset V$ is the space of R-R ground states in the $20$ $\CN=(4,4)$ representations $(\frac{1}{4},0;\frac{1}{4},0)$. The basic idea of \cite{Gaberdiel:2011fg} is to embed $L_G(-1)$ into the Leech lattice $\Lambda$, whose automorphism group is the Conway group $Co_0$. In this way, for a given $K3$ model, it was possible to characterize  all groups $G$ acting trivially on $\Pi$ and preserving the $\CN=(4,4)$ superconformal algebra as the subgroups of the Conway group  $Co_1 \subset Co_0 = Aut(\Lambda)$ fixing pointwise a sublattice of $\Lambda$ of rank at least $4$.

This characterization of the symmetries of $K3$ NLSMs establishes a very close connection between these models and the Conway moonshine module $V^{f \natural}$ \cite{Frenkel:1988flm,Duncan:2006}, which is notoriously the unique $c \leq 12$ holomorphic SCFT without continuous automorphisms, and whose automorphism group is $Co_0$.

In recent years, the concept of symmetry in a Quantum Field Theory (QFT) has been found to be strictly related to the one of topological defects, which are invariant under continuous deformations as
long as the deformations do not move the inhomogeneities past other defects or field operator
insertions. Topological defects naturally generalize the notion of symmetry to include higher-form symmetries \cite{Gaiotto:2014kfa}  and non-invertible symmetries.  The prototypical example of the latter being the ``Kramers-Wannier'' duality defect in the critical Ising model \cite{Frohlich:2009gb, Frohlich:2006ch, Frohlich:2004ef, Kramers:1941kn}.  In the context of 2-dimensional rational CFTs, topological defect lines (TDL) are seen to realize the well-known mathematical structure of fusion categories \cite{Bhardwaj:2017xup,Chang_2019,Thorngren:2019iar,Thorngren:2021yso}. Whereas TDLs have been widely studied in rational CFTs \cite{Cardy:1986gw,Fuchs:2002cm, Fuchs:2003id, Fuchs:2004dz, Fuchs:2004xi,Oshikawa:1996dj, Oshikawa:1996ww, Petkova:2000ip, Verlinde:1988sn, Zuber:1986ng}, thanks to the fact that TDLs preserving  rational chiral and anti-chiral algebras satisfy a Cardy-like condition and therefore depend on finitely many complex parameters, which determine how the defect acts on the (finitely  many) primary operators,  for non-rational CFTs an analog approach would involve solving for an infinite number of parameters, which is generally unfeasible.  Only for a few torus models a full classification has been attempted\cite{Fuchs:2007tx,Bachas:2012bj,Thorngren:2021yso}. $K3$ models are another natural setup of non-rational CFT where the large amount of space-time
and worldsheet supersymmetries, which allows for a precice description of the geometry of the moduli space, the elliptic genus (which is the same for every K3 model), the spectrum of short $\CN=(4,4)$ representations, and even the finite groups of symmetries at each point in the moduli space \cite{Aspinwall:1996mn,Gaberdiel:2011fg,Nahm:1999ps},  makes a partial classification possible.

In this note, in particular, we review recent progress on TDLs in $K3$ models presented in \cite{Angius:2024evd}, based on a generalization of the lattice-theoretic approach of \cite{Gaberdiel:2011fg}. This is part of an ongoing effort in understanding the still somewhat obscure relationship between $K3$ models and $V^{f\natural}$ (see \cite{Volpato:2024goy,Angius:2025xxx,Angius:2025zlm}). 

In section \eqref{sec:GeneralFacts} we concisely review known facts about $K3$ models and $V^{f\natural}$, with a focus on symmetries, which can be seen as the subcategory of invertible TDLs. In section \eqref{sec:TDLs} we review some basic facts about topological defects in two dimensional CFTs, and fix the notation. In section \eqref{Section:4} we present the main general facts that can be derived from the lattice-theoretic approach. In particular, we argue that, while for certain K3 models infinitely many simple defects, and even a continuum,
can occur, at generic points in the moduli space the category is actually trivial, i.e. it is generated
by the identity defect. If a $K3$ model is at the attractor point for some
BPS configuration of D-branes, then all topological defects have integral quantum dimension.
We also present a conjecture that a continuum of topological defects arises if and only if the $K3$ model is
a (possibly generalized) orbifold of a torus model. We stress that the presented results are proven on a physics level of rigor, as they are based on various assumptions about K3 models and boundary states that are not mathematically rigorous. In section  \ref{sec:Example} we discuss a particular $K3$ model, namely the model $\calC_{\GTVW}$ with 
`the largest symmetry group', which was first discussed in \cite{Gaberdiel:2013psa}, and to which we refer  as the GTVW model, and for which the relationship to $V^{f\natural}$ is particularly explicit \cite{Creutzig:2017fuk,Taormina:2017zlm}. Finally, in section \ref{sec:conclusions}, we summarize our discussion and point out connections to lines of ongoing research.

\section{$K3$ models and $V^{f\natural}$}
\label{sec:GeneralFacts}
In this section, we introduce $K3$ non-linear sigma models and the notation that will be used in the following. We also discuss the relationship with the Conway's module $V^{f\natural}$. 
\subsection{$K3$ non-linear sigma models}
\label{sec:K3NLSM}
Non-linear sigma models (NLSMs) on $K3$ are two-dimensional  theories with $\mathcal{N}=(4,4)$ superconformal symmetry and central charge $(c, \bar{c})=(6,6)$. These theories play a prominent role in physics, as they describe the worldsheet dynamics of type $II$ superstrings compactified on the simplest class of compact Calabi–Yau manifolds: the $K3$ surfaces. Notably, type $IIA$ compactification on $\mathcal{M}^{1,3} \times T^2 \times K3$ provided one of the earliest concrete examples of the AdS/CFT correspondence and offered the first successful framework for a microscopic derivation of the Bekenstein–Hawking entropy on five-dimensional supersymmetric black holes \cite{Strominger:1996sh}. From a mathematical perspective, $K3$ sigma models have attracted significant attention due to their deep connections with sporadic simple groups and their associated moonshine phenomena. Remarkably, the only other known class of unitary two-dimensional theories with the same superconformal symmetry are those with target space $\mathbb{T}^4$. These two classes are sharply distinguished by their elliptic genera:
\begin{equation}
    \phi (\tau, z) = Tr_{\mathcal{H}_RR} \left( q^{L_0- \frac{c}{24}} \bar{q}^{\bar{L}_0 - \frac{\bar{c}}{24}} (-1)^{F + \bar{F}} y^{J_0}\right)  \, \,  = \, \,  \begin{cases} \, \, \, \, 
         0  \quad \quad & \mathbb{T}^4 \\ \, \, \, \,  8 \sum_{i=1}^4 \frac{\theta_i (\tau, z)^2}{\theta_i (\tau, 0)^2} \quad \quad  &K3,
    \end{cases}
    \label{K3:Elliptic_genus}
\end{equation}
where $\theta_i(\tau,z)$ are the standard Jacobi theta functions. Such a functions are invariant under exactly marginal deformations that preserve the $\mathcal{N}=(4,4)$ superconformal algebra and span the moduli spaces $\mathcal{M}_{\mathbb{T}^4}$ and $\mathcal{M}_{K3}$. In the latter case, the $80-$dimensional moduli space of non-linear sigma models on $K3$ takes the form:
\begin{equation}
    \mathcal{M}_{K3}= O \left( \Gamma^{4,20} \right) \backslash O(4,20) /\left( O(4) \times O(20)\right),
    \label{K3:moduli_space}
\end{equation}
where the coset $O(4,20)/\left( O(4) \times O(20) \right)$ is a Grassmannian that parametrizes the choice of a positive-definite four-dimensional subspace $\Pi$ in $\mathbb{R}^{4,20}$. The subspace $\Pi \subset \mathbb{R}^{4,20}$ encodes the choice of a Ricci-flat metric on the target $K3$ surface along with a background $B$-field. The discrete group $O(\Gamma^{4,20})$  denotes the automorphism group of the even unimodular lattice $\Gamma^{4,20} \subset \mathbb{R}^{4,20}$ of signature $(4,20)$.  The ambient space $\mathbb{R}^{4,20}$ can be identified with the even cohomology group $H^{even} (K3, \mathbb{R})$ of $K3$, while the lattice $\Gamma^{4,20}$ corresponds to the even integral homology $H_{even}(K3, \mathbb{Z})$ embedded in $\mathbb{R}^{4,20}$ via Poincar\'e duality. \\
To interpret the moduli space \eqref{K3:moduli_space} from a physical perspective, it is necessary to consider the structure of the space of states of this class of NLSMs. Both the holomorphic and anti-holomorphic sectors realize the small $\mathcal{N}=4$ superconformal algebra at central charge $c=6$. This algebra consists of a stress-energy tensor and an affine $\hat{su}(2)_1$ current algebra in the bosonic sector, and four supercurrents in the fermionic sector. There exist two classes of unitary irreducible representations of this algebra, defined in both the Neveu-Schwarz (NS) and Ramond (R) sectors, and related by spectral flow. These representations are labeled by pairs $(h,q)$, where $h$ is the conformal weight and $q \in \mathbb{Z}/2$ is the eigenvalue with respect to the zero mode of the Cartan current $J^3_0$ of $\hat{su}(2)_1$ of the highest states of the representation. In particular, there exists a set of two short (BPS) representations in the NS sector labeled by the pairs $(0,0)$ and $\left( \frac{1}{2}, \frac{1}{2}\right)$, which are mapped under spectral flow to the R-sector representations $\left( \frac{1}{4}, \frac{1}{2} \right)$ and $\left( \frac{1}{4}, 0 \right)$, respectively. The remaining class of long (non-BPS)  form an infinite family labeled by the pairs $\left(h , 0 \right)$, with $h > 0$, in the NS sector and their R-sector counterparts $\left(h, \frac{1}{2} \right)$, with $h > \frac{1}{4}$. In the full $\mathcal{N}=(4,4)$ superconformal theory, one defines as $\frac{1}{2}$-BPS those representations that are BPS in both left- and right-moving sectors (i.e. short-short representations). For any K3 sigma model, the $\frac{1}{2}$-BPS content in the NS-NS sector comprises the vacuum representation $(0,0; 0,0)$, corresponding to the identity operator, and twenty additional representations of the form $\left(\frac{1}{2}, \frac{1}{2}; \frac{1}{2}, \frac{1}{2}\right)$, associated with the harmonic $(1,1)$ forms on the K3 surface.  Under spectral flow, the vacuum maps to the four R-R ground states $\left( \frac{1}{4}, \frac{1}{2}; \frac{1}{4}, \frac{1}{2} \right)$, which generate the spectral flow automorphisms of the algebra, while the other twenty representations map to twenty additional R-R ground states $\left( \frac{1}{4}, 0; \frac{1}{4}, 0 \right)$. Beyond this topological ($\frac{1}{2}$-BPS) sector, each $K3$ sigma model contains an infinite spectrum of mixed BPS and non-BPS representations (short-long, long-short, and long-long), whose content depends on the specific model. In contrast to supersymmetric sigma models on target $\mathbb{T}^4$, $K3$ models do not admit free fermionic representations such as $\left( \frac{1}{2},\frac{1}{2}; 0,0 \right)$ and $\left( 0,0; \frac{1}{2}, \frac{1}{2} \right)$. Within this framework, the lattice $\Gamma^{4,20}$ appearing in \eqref{K3:moduli_space}  can be interpreted as the lattice of R-R D-brane charges, associated with the $24$ R-R ground states. This lattice embeds into the real space of R-R ground states $\mathbb{R}^{4,20} \simeq \Gamma^{4,20} \otimes \mathbb{R}$,  and the group $O\left( \Gamma^{4,20} \right)$ plays the role of the T-duality group. The positive-definite subspace $\Pi$ is naturally identified with the real four-dimensional subspace that supports the spectral flow representations. \\
As noted above, there is a rich geometric interplay between $K3$ symmetries and certain sporadic simple groups.  The first insight in this direction is the  Mukai’s theorem \cite{Mukai:1988, Kondo:1988}, which provides a complete classification of geometrical symmetries of K3 surfaces in terms of subgroups of the Mathieu group $\mathbb{M}_{23}$. Interest in this connection deepened following the work of Eguchi, Ooguri and Tachikawa \cite{Eguchi:2010ej} who observed that the elliptic genus of $K3$ encodes an infinite-dimensional graded representation of $\mathbb{M}_{24}$, the larger Mathieu group containing $\mathbb{M}_{23}$ as a maximal subgroup. Since the elliptic genus is invariant across the moduli space, this observation suggests the existence of an hidden $\mathbb{M}_{24}$ symmetry associated with the corresponding $K3$ sigma-models, encapsulating both geometric ($\mathbb{M}_{23}$) that non-geometric symmetries. This insight motivated the generalization of Mukai’s theorem to full NLSM symmetries in the work of  Gaberdiel, Hohenegger, and Volpato \cite{Gaberdiel:2011fg}. In particular, given a $K3$ NLSM  $\mathcal{C}_{\Pi}$, associated with the choice of a positive-definite four-plane $\Pi \subset \mathbb{R}^{4,20}$, i.e. a point in the Grassmannian space, they proved that: 
\begin{itemize}
    \item[\textit{(i)}] The corresponding symmetry group $G_{\Pi}$ of elements preserving both the $\mathcal{N}=(4,4)$  superconformal algebra and spectral flow generators is always isomorphic to the subgroup $Stab(\Pi)$ of the lattice automorphisms group $Aut(\Gamma^{4,20}) \simeq O(\Gamma^{4,20})$ fixing the subspace $\Pi \subset \mathbb{R}^{4,20}$ pointwise:
    \begin{equation}
        G_{\Pi} \simeq Stab (\Pi) = \left\lbrace g \in O (\Gamma^{4,20}) \, \vert \, \,  g_{\vert_{\Pi}} = \text{id} \right\rbrace \subset O (\Gamma^{4,20});
    \end{equation}
    \item[\textit{(ii)}] The group $Stab (\Pi)$ is always isomorphic to a subgroup of the Conway group $Co_1 \subset Co_0 = Aut(\Lambda)$ that fixes pointwise a sublattice of the Leech lattice $\Lambda$ of rank at least 4. 
\end{itemize}
The proof of the theorem proceeds as follows. Since every symmetry $g \in G_{\Pi}$ preserves the $\mathcal{N}=(4,4)$ superconformal algebra,  it also preserves the representation of states. In particular, $g$ maps $\frac{1}{2}$-BPS boundary states into $\frac{1}{2}$-BPS boundary states,  thereby inducing an automorphism on the lattice of R–R D‑brane charges:
\begin{equation}
    \rho \, \, : \, \, G_{\Pi} \, \, \mapsto \, \, Stab (\Pi) \subset Aut (\Gamma^{4,20}).
\end{equation}
This map is actually an isomorphism. By definition, every element of $Stab(\Pi)$ fixes the four‑plane $\Pi$ and hence induces a self‑duality of $\mathcal{C}_{\Pi}$. Therefore each such element arises as some symmetry $g \in G_{\Pi}$, so $\rho$ is onto. Injectivity follows from the fact that the elements of the kernel $ker(\rho) \subset G_{\Pi}$ commute with the spectral flow operators and define a trivial action on all the R-R ground states then they have to act trivially also  in the 80 NS-NS exactly marginal operators spanning the moduli space. Since it can be shown that the kernel is trivial for a specific K3 model, it has to be trivial everywhere in $\mathcal{M}_{K3}$, then the map is injective. \\
Thus, the problem of classifying symmetry groups of $K3$ NLSMs reduces to classifying subgroups of $Co_0$ fixing a sublattice of rank at least $4$ in the lattice $\Lambda$. As a result, any such symmetry group $G$ falls into one of four types:
\begin{itemize}
    \item[(1)] $G=G'.G''$, where $G'$ is a subgroup of $\mathbb{Z}_2^{11}$, and $G$ is a subgroup of $\mathbb{M}_{24}$ with at least four orbits when acting as a permutation on $\left\lbrace 1,2,...,24 \right\rbrace$;
    \item[(2)] $G=5^{1+2}.\mathbb{Z}_4$;
    \item[(3)] $G=\mathbb{Z}_3^4.A_6$;
    \item[(4)] $G=3^{1+4}.\mathbb{Z}_2. G''$, where $G''$ is either trivial, $\mathbb{Z}_2$, $\mathbb{Z}_2^2$ or $\mathbb{Z}_4$. 
\end{itemize}
Note that, except in case $(1)$ when $G'$ is trivial, these groups are not contained in $\mathbb{M}_{24}$; rather, they lie properly within $Co_1 \subset Co_0$  extending the role of Mathieu symmetries in $K3$ sigma models.

\subsection{The Conway module $V^{f \natural}$}
The Conway moonshine module $V^{f \natural}$ \cite{ Duncan:2006} is one of exactly three holomorphic, two-dimensional $\mathcal{N}=1$ superconformal field theories (SCFTs) with central charge $c=12$. The other two are:
\begin{itemize}
\item $F_{24}$, the theory generated by $24$ free chiral (Majorana–Weyl) fermions; 
\item $V^{f E_8}$, the $\mathcal{N}=1$ supersymmetrization of the lattice vertex operator algebra (VOA) $V^{E_8}$ based on the lattice $E_8$. 
\end{itemize}
Among these, $V^{f \natural}$ is uniquely distinguished by its minimal conformal weight $\frac{3}{2}$ for fermions; in this sense, $V^{f \natural}$ is considered the supersymmetric analogue of the Frenkel-Lepowsky-Meurman Monster module $V^{\natural}$ \cite{Frenkel:1988flm}, which is conjecturally the unique holomorphic bosonic VOA with $c = 24$ and no states of conformal weight $1$.\\
There are several equivalent constructions of this theory. The simplest one introduces $12$ holomorphic free bosons $X^i (z)$, $i=1,2,...,12$, which generate $12$ holomorphic $u(1)$ currents $i \partial X^i(z)= \sum_{n=- \infty}^{+ \infty} \alpha_{n}^i z^{-n-1}$ of weight $1$ and standard OPE
\begin{equation}
    i \partial X^i (z) \, i \partial X^j (w) = - \frac{\delta^{ij}}{(z-w)^2}.
\end{equation}
One then includes the full set of vertex operators
\begin{equation}
    \mathcal{V}_k (z) = c_k : e^{i k \cdot X(z)}:
\end{equation}
with momenta $k \equiv \,  {}^t(k^1,k^2,...,k^{12})$ taking values in the odd unimodular lattice
\begin{equation}
    D_{12}^+= \left\lbrace \frac{1}{2} (x_1, x_2, ..., x_{12}) \in \left( \frac{1}{2} \mathbb{Z}\right)^{12} \, \vert \, x_i \equiv x_j \, \,  \text{mod} \, \, 2 \, , \sum x_i \in 4 \mathbb{Z} \right\rbrace \simeq D_{12} \, \cup (D_{12} +s)
\end{equation}
containing the root lattice $D_{12}$ of the $so(24)$ Lie algebra and its spinor-class translate $(D_{12}+s)$, where $s= \left( \frac{1}{2},  \frac{1}{2}, ..., \frac{1}{2} \right) \in \mathbb{R}^{12}$. \\
The stress-energy tensor of the theory has the standard form $T(z)= \frac{1}{2} \sum_{i=1}^{12} : \partial X^i  \partial X^i : (z)$ and central charge $12$. Under the state–operator correspondence, the operators $\mathcal{V}_k$ correspond to the states $\vert k \rangle $ with charges $(k^1,k^2,...,k^{12})$ with respect to the $u(1)^{12}$ currents
\begin{equation}
    \alpha^i_0 \vert k \rangle = k^i \vert k \rangle
\end{equation}
and conformal weight $k^2/2$. In particular, states with even $k^2$ are bosonic, while those with odd $k^2$ are fermionic. As stated above, since the lattice $D_{12}^+$ contains no vectors of squared length $1$, the minimal conformal weight for fermionic states is  $3/2$, and precisely $2^{11}=2048$  such states exist at that level. The bosonic subalgebra of $V^{f \natural}$ is the affine Kac-Moody algebra $\hat{so}(24)_1$ generated by the $12$ abelian $u(1)$ currents and the bosonic vertex operators with momenta in the root lattice $D_{12}$. In terms of irreducible modules of this algebra, the theory decomposed as
\begin{equation}
    V^{f \natural} = V_0 \, \oplus \, V_s,
\end{equation}
where $V_0$ is the vacuum module, and $V_s$ is the spinorial module corresponding to the non-trivial coset $s+ D_{12}$. Physically, $V^{f \natural}$  realizes only the Neveu–Schwarz (NS) sector of the full superconformal theory. The Ramond (R) sector is recovered by considering the remaining two irreducible $\hat{so}(24)_1$ modules
\begin{equation}
    V^{f \natural}_{tw} = V_v \, \oplus \, V_c
\end{equation}
associated to the non-trivial cosets $v + D_{12}$ and $c+ D_{12}$ of $D_{12}^{\ast}/D_{12}$, where $v= \left( 1, 0,0,...,0\right) \in \mathbb{R}^{12}$ and $c= \left( - \frac{1}{2}, \frac{1}{2}, ..., \frac{1}{2} \right) \in \mathbb{R}^{12}$, labeling the vector and conjugate spinor representations of $so(24)$, respectively. The subscript in $V^{f \natural}_{tw}$ indicates that this module is obtained as the canonical $\mathbb{Z}_2$-twist of $V^{f \natural}$ by the fermion number symmetry $  (-1)^F $, which acts trivially on $V_0$ and by $-1$ on $V_s$. \\
The full automorphism group of the holomorphic SCFT $V^{f \natural}$ is $Spin(24)$. However, the group of faithful automorphisms is $Spin(24)/ \mathbb{Z}_2$, which $\mathbb{Z}_2$ generator is the lift to $Spin(24)$ of $-1$ in $SO(24)$ and acts non-trivially on $V^{f \natural}_{tw}$. For this reason, we consider the full group $Spin(24)$  in what follows. \\
There are two other constructions of the theory $V^{f \natural}$, each highlighting its connection with one of the other holomorphic SCFTs with $c=12$. The first arises from the $\mathbb{Z}_2$ orbifold of the free fermion theory $F_{24}$
\begin{equation}
    V^{f \natural} = \, F_{24}  \, / \, \mathbb{Z}_2,
\end{equation}
in which the symmetry $\mathbb{Z}_2$ acts by reversing the sign of all the $24$ Majorana-Weyl fermions generating the theory.\\
The second construction relates $V^{f \natural}$ with $V^{f E_{8}}$. This latter theory is obtained extending the VOA $V^{E_8}$ of $8$ chiral free bosons $X^i(z)$, $i=1,2,...,8$, compactified on the lattice $E_8$, with the addition of $8$ chiral free fermions $\psi^i(z)$, $i=1,2,...,8$:
\begin{equation}
    V^{f E_8} = \, V^{E_8} \, \otimes \, F_8.
\end{equation}
The $\mathcal{N}=1$ supercurrent is constructed  explicitly using these fields via $ \tau(z) = \sum_{i=1}^8 \psi^i X^i $. The theory owns a $\mathbb{Z}_2$ symmetry acting with a sign $-1$ on bosons and fermions and leaving the supercurrent invariant. Gauging this symmetry yields a theory isomorphic to  $V^{f \natural}$ \cite{Duncan:2006}
\begin{equation}
    V^{f \natural} \, \simeq \, V^{f E_8} \, / \, \mathbb{Z}_2.
\end{equation}
Notably, since the symmetry $\mathbb{Z}_2$ preserve the supercurrent, this one is constantly transported in $V^{f \natural}$. This latter presentation is particularly well-suited for constructing topological defect lines in $V^{f \natural}$ that manifestly preserve the supercurrent, see \cite{Angius:2025xxx} for explicit applications. \\
The $\mathcal{N}=1$ supercurrent in $V^{f \natural}$ is unique up to the action of $Spin(24)$ and the subgroup of automorphisms preserving a chosen $\tau(z)$ is isomorphic to the finite group of Conway:
\begin{equation}
    Aut_{\tau} (V^{f \natural}) = \left\lbrace g \in Aut (V^{f \natural}) \, \vert \, g (\tau)=\tau  \right\rbrace \simeq Co_0.
\end{equation}

\subsection{The connection between $K3$ models and $V^{f \natural}$}
The appearance of the Conway group in both K3 NLSMs and $V^{f \natural}$ provides the basis of the mutual connection among the two theories. The first explicit link was established in \cite{Duncan:2015xoa} by observing that every discrete symmetry group of a K3 sigma model, which preserves the small $\mathcal{N}=(4,4)$ superconformal algebra and spectral flow, embeds into the Conway group $Co_0$, the automorphism group of the Leech lattice $\Lambda$, the unique even unimodular lattice in $24$ dimensions without roots. \\
Concretely, for each $K3$ model its symmetry group $G_{\Pi^{\natural}}$ is a subgroup of $ Co_0$ fixing a real four‑plane $\Pi^{\natural} \subset \mathbb{R}^{24}$ in the $\mathbf{24}$ vectorial representation of $SO(24) \supset Co_0$, and conversely any choice of such a group determines at least a K3 model with symmetry exactly $G_{\Pi^{\natural}}$. Crucially, this correspondence is not merely group‑theoretic: the action of each $g \in G_{\Pi^{\natural}}$ on the topological sector of the K3 model coincides with the action of the corresponding symmetry in $Co_0$ on the Ramond ground states sector ${}^{\mathbb{R}}V^{f \natural}_{tw} (1/2)$ of the twisted module  of $V^{f \natural}$. This sector consists of $24$ states of minimal conformal weight $1/2$ spanning a real vector space isomorphic to $\mathbb{R}^{24}$.\\
Seeing $V^{f \natural} \oplus V^{f \natural}_{tw}$ in terms of irreducible modules of $\hat{so}(24)_1$, fixing the four-plane $\Pi^{\natural} \subset \mathbb{R}^{24}$ means to single out a subalgebra $so(4)_1 \simeq su(2)_1 \oplus su(2)_1 \subset so(24)_1$. Now, by choosing one $su(2)_1$ factor together with the unique $\mathcal{N}=1$ supercurrent $\tau (z)$, one embeds a copy of the small $\mathcal{N}=4$ algebra at $c=6$ inside the $\mathcal{N}=1$ with $c=12$ super-algebra \cite{Cheng:2016org}. This makes possible the definition of a graded partition function in $V^{f \natural}$
\begin{equation}
    \phi \left( V^{f \natural}, \tau, z \right) = Tr_{V^{f \natural}_{tw}} \left( (-1)^F q^{L_0 - 1/2} y^{J^3_0} \right),
\end{equation}
where $J_0^3$ is the zero mode of the $su(2)_1$ Cartan generator, which coincides exactly with the elliptic genus \eqref{K3:Elliptic_genus} of $K3$ models. Furthermore, for each $g \in G_{\Pi^{\natural}}$ in the $K3$ model $\mathcal{C}$ one can define the twined elliptic genus
\begin{equation}
    \phi^g \left( \mathcal{C}, \tau, z \right) = Tr_{RR} \left( \hat{g} (-1)^F q^{L_0 - 1/4} \bar{q}^{\bar{L}_0 -1/4} y^{J_0^3}\right)
    \label{K3:twining_genus}
\end{equation}
which only depend on the conjugacy class of $g \in O(\Gamma^{4,20})$. Similarly, on the $V^{f \natural}$ side, for each $g \in Co_0$ preserving a four-plane one can define the graded twined partition function
\begin{equation}
    \phi^g \left( V^{f \natural}, \tau, z \right) = Tr_{V_{tw}^{f \natural}} \left( \hat{g} (-1)^F q^{L_0 - 1/2}  y^{J_0^3} \right)
    \label{Vfnat:twining_part_funct}
\end{equation}
which only depend on the conjugacy class of $g \in Co_0$. Although there is not a one-by-one correspondence among the conjugacy classes of $O(\Gamma^{4,20})$ and $Co_0$, almost all the twining genera of $K3$ NLSMs are reproduced by a corresponding twining partition function in $V^{f \natural}$ of the form \eqref{Vfnat:twining_part_funct}\footnote{The few exceptions are discussed in \cite{Cheng:2016org}.} These parallel results exemplify two intertwining moonshine phenomena: on one hand, the graded traces of $Co_0$ on $V^{f \natural}$ and its twisted module are genus‑zero principal moduli, establishing Conway moonshine \cite{Duncan:2015xoa}; on the other hand, the elliptic genus and its twined versions for K3 realize Mathieu and Umbral moonshine, connecting K3 NLSMs to sporadic groups such as $\mathbb{M}_{24}$. 
Despite the evident parallel between the symmetry structures of $K3$ NLSMs and the Conway module, an explicit construction of their correspondence remains known only in a single, exceptionally symmetric K3 model, the so‑called Gaberdiel-Taormina-Volpato-Wendland model $\mathcal{C}_{GTVW}$, studied in \cite{Gaberdiel:2013psa}, which possesses the largest finite symmetry group among all $K3$ sigma-models. In that case, one realizes the link by a reflection operation that sends each field in $\mathcal{C}_{GTVW}$ of conformal weight $(h, \bar{h})$ to a purely holomorphic field in $V^{f \natural}$ of weight $h+\bar{h}$, see \cite{Taormina:2017zlm, Creutzig:2017fuk} for a detailed discussion. The absence of a comparable mechanism for general points in the $K3$ moduli space continues to motivate ongoing investigation into this problem. In light of recent progress on generalized notions of symmetry in quantum field theory, it is natural to ask whether the Conway–$K3$ correspondence, so far established for ordinary (0‑form) symmetries, extends to the realm of generalized (higher‑form or non‑invertible) symmetries.

\section{Topological defects as generalized symmetries in CFTs}
\label{sec:TDLs}
In this section we provide a concise overview of the main definitions and properties of topological defects in generic two‑dimensional conformal field theories (CFTs), discussing the principal classification techniques and highlighting the main open problems.\\
It is well-established that in generic Quantum Field Theories (QFTs) defined on a $d$-dimensional spacetime $\mathcal{M}_d$, extended objects associated with non-local operators can exist alongside local operators.  Such objects are called \textit{defects}, and the corresponding operators, supported on submanifolds $\mathcal{M}^{d-q}$ of spacetime with positive codimension, i.e. $q < d$, are denoted with the notation $\hat{D} \left( \mathcal{M}^{d-q} \right)$. Operators of this type are called \textit{topological} if arbitrary small deformations of their support manifold, which do not cross other operators insertions, leave all the correlators invariant. \\
The simplest examples of topological defects in any QFT arise from ordinary global symmetries. The construction is the following. Given a continuous global symmetry group $G$, Noether's theorem associates to each generator $T_{(a)}$ ($a=1,2,\dots,\dim G$) a conserved $(d-1)$-form current $\ast j_{(a)}$ satisfying $d(\ast j_{(a)}) = 0$. Any element $g$ in the identity component of $G$ can be expressed in terms of a set of parameters $\alpha^{(a)}$ as
\begin{equation}
    g = e^{i \alpha^{(a)} T_{(a)}}.
\end{equation}
By exponentiating the integrated current for these parameters, for each element $g \in G$ one can construct an operator supported on a closed oriented codimension-1 submanifold of spacetime:
\begin{equation}
    \hat{D}_g  \left( \mathcal{M}^{d-1}\right) = \exp\left(i \alpha^{(a)} \oint_{\mathcal{M}^{d-1}} \ast j_{(a)} \right).
    \label{top_defect:standard_symm}
\end{equation}
Placing two such defect operators, associated to $g$ and $g' \in G$, on the same support manifold allows one to define a generalized OPE of the corresponding non‑local operators that exactly reproduces the group multiplication law of $G$:
\begin{equation}
    \hat{D}_g \left( \mathcal{M}^{d-1} \right) \times \hat{D}_{g'} \left( \mathcal{M}^{d-1} \right) = \hat{D}_{g''} \left( \mathcal{M}^{d-1} \right) \quad \quad \text{with } \, \, g''=g \cdot g' \in G.
    \label{OPEs:group_law}
\end{equation}
Under small deformations of the support manifold $\mathcal{M}^{d-1} \to \mathcal{M}'^{d-1}$, Figure \ref{fig1}, that avoid other operators insertions, Stokes' theorem shows that operators defined as in \eqref{top_defect:standard_symm} are topological:
{\small{
\begin{eqnarray}
   \hat{D}_g \left( \mathcal{M}^{d-1} \right) \times \hat{D}_g^\dagger \left( \mathcal{M}'^{d-1} \right) 
   &=& \exp\left[i \alpha^{(a)} \left(  \oint_{\mathcal{M}^{d-1}}  \ast j_{(a)}  -  \oint_{\mathcal{M}'^{d-1}} \ast j_{(a)} \right)\right] \nonumber\\
   &=& \exp\left[i \alpha^{(a)} \int_{\mathcal{N}^{d}} d\ast j_{(a)} \right] 
   = \mathbb{I},
\end{eqnarray}
}}
where $ \mathcal{N}^d$ is the region bounded by $\mathcal{M}^{d-1}- \mathcal{M}'^{d-1}$.
\begin{figure}[h!]
    \centering
    \includegraphics[width=0.6\linewidth]{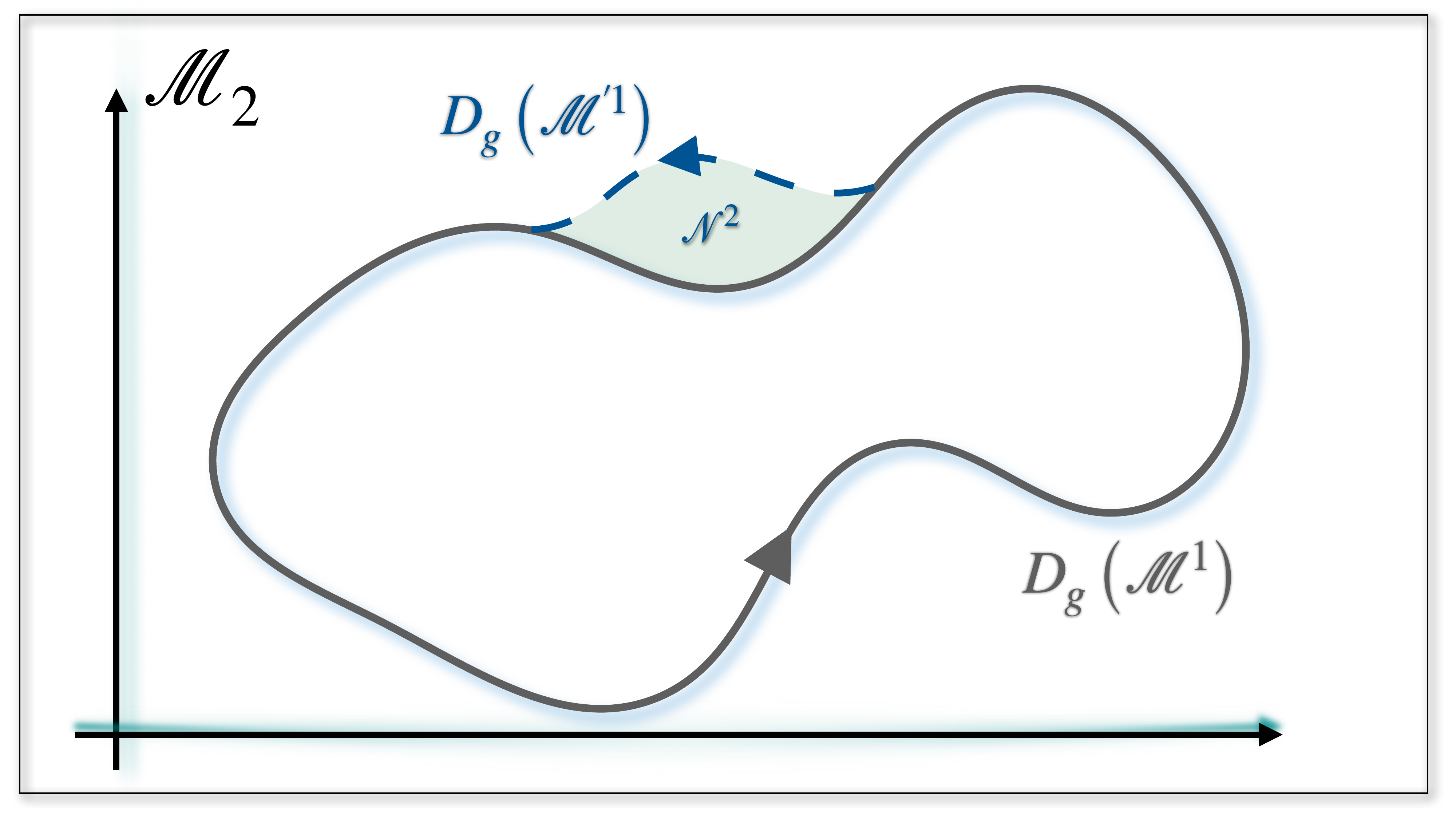}
    \caption{Small deformation $\mathcal{M}^1 \mapsto \mathcal{M}'^1$ of the support manifold of the extended operator $\hat{D}_g$ on the two-dimensional spacetime $\mathcal{M}_2$.}
    \label{fig1}
\end{figure}

During the last few decades, this reformulation of ordinary symmetries as topological defects supported on codimension-$1$ slices of spacetime has inspired a wide array of generalizations of the standard symmetry paradigm. 
  The first such extension arises by relaxing the requirement that defects be codimension-$1$ and instead permitting defects of arbitrary codimension. This leads directly to the notion of $p$-form global symmetries \cite{Gaiotto:2018ypj}, whose symmetry defects are supported on codimension-$(p+1)$ submanifolds of the ambient spacetime and act on non-local charged operators supported on $p$-dimensional submanifolds. In this language, the familiar ordinary symmetries are precisely the case $p=0$, for this reason they are called $0$-form symmetries. A second, independent generalization allows defects whose OPEs no longer close under a group law, as in \eqref{OPEs:group_law}, but instead obey to more general fusion rules. Concretely, if $D_a$ and $D_b$ are two topological defects supported on a $(d-q)$-dimensional submanifold, in the limit in which their support manifold overlap, the generalized OPE between their corresponding operators define a \textit{fusion algebra} of the form
\begin{equation}
    \hat{D}_a \left( \mathcal{M}^{d-q}\right) \times \hat{D}_b \left( \mathcal{M}^{d-q}\right) = \sum_c N^{c}_{ab} \hat{D}_c \left( \mathcal{M}^{d-q}\right),
\end{equation}
where the right side is the sum over finitely many extended operators with respect to the integral fusion coefficients $N^c_{ab}$. These generalized fusion algebras underlie the modern theory of \textit{non‑invertible} categorical symmetries in QFT.

\subsection{Generalities on TDLs in 2D CFTs}
Since the primary aim of this article is the the study of topological defects in K3 NLSMs, in the rest of this section we will focus our discussion on general aspects of topological defects in $2D$ CFTs, though many of these concepts extend naturally to general and higher dimensional QFTs. \\
In two dimensions, extended operators are supported on one-dimensional submanifolds $\gamma$ of spacetime. We refer to such operators as Topological Defect Lines (TDLs), denoted $\mathcal{L}(\gamma)$, with $\hat{\mathcal{L}}$ representing the corresponding operator.\\
Let $S= \left\lbrace \mathcal{L}_a \right\rbrace$  be the set of TDLs in a $2D$ QFT $\mathcal{Q}$ defined on the spacetime $\Sigma$. These defects satisfy the following fundamental properties:
\begin{itemize}
    \item The set $S$ always includes a distinguished trivial defect $\mathcal{I}$, corresponding to the identity operator. It is conventionally depicted as a dotted line, see Figure \ref{fig2}-(a).
    \item A support line may be closed or open. An open line can either extend to infinity or terminate on local operators. In the latter case, its allowed starting-point operators form a vector space $\mathcal{H}_{\mathcal{L}}$, called the $\mathcal{L}$-twisted sector, whose elements are known as \textit{defect operators}, see Figure \ref{fig2}-(b).
    \begin{figure}[h!]
        \centering
        \includegraphics[width=0.6\linewidth]{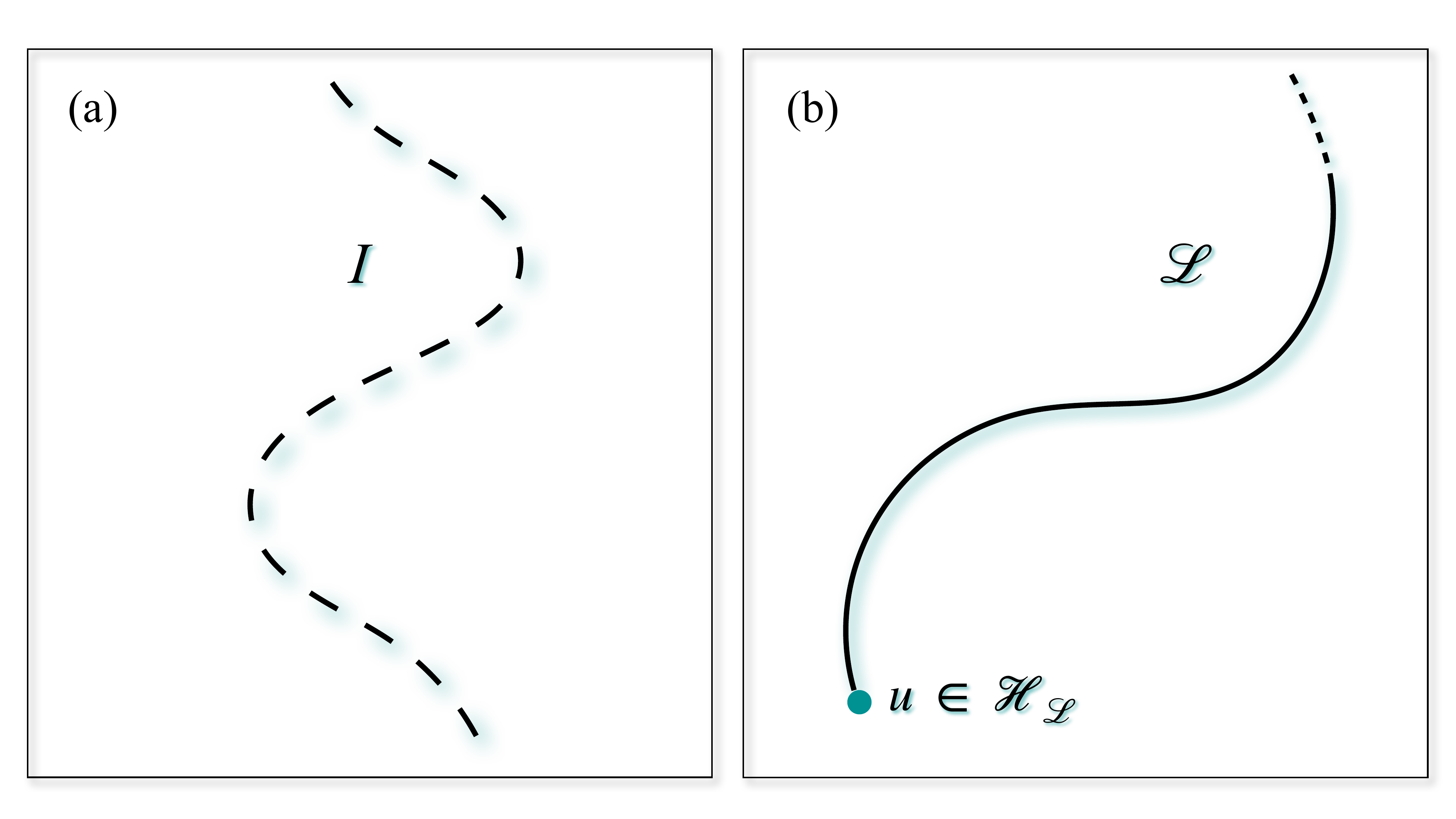}
        \caption{(a) Trivial defect associated with the identity operator.  (b) An endable TDL $\mathcal{L}$, terminating on the defect operator $u \in \mathcal{H}_{\mathcal{L}}$.}
        \label{fig2}
    \end{figure}
    
    \item All TDLs are oriented. The set $S$ admits an involution $\mathcal{L}  \mapsto \mathcal{L}^{\ast}$ that reverses the orientation $\gamma \mapsto \bar{\gamma}$ of the support lines:
    \begin{equation}
        \mathcal{L}^{\ast} (\gamma) \cong \mathcal{L} (\bar{\gamma}).
    \end{equation}
    The twisted sector of the reversed defect satisfies $\mathcal{H}_{\mathcal{L}}^{\ast} \cong \mathcal{H}_{\mathcal{L}^{\ast}}$, where $\mathcal{H}_{\mathcal{L}^{\ast}}$ can be interpreted as the vector space of end-point operators for the endable line $\mathcal{L}$  . At the operator level, this relates to the adjoint via:
    \begin{equation}
        \hat{\mathcal{L}}^{\dagger} = \theta \hat{\mathcal{L}}^{\ast} \theta
    \end{equation}
    where $\theta= \theta^{-1}$ is the CPT anti-linear involution.
    \item TDLs act on the local operators. A pictorial representation of this action is realized by encircling the local operator $\hat{\mathcal{O}}(z)$ with a closed line $\gamma$ supporting the defect $\mathcal{L} \in S$. Since the defect is topological, one can shrink its support over the local operator $\hat{\mathcal{O}}(z)$ and get a new operator $\hat{\mathcal{LO}}(z)$, Figure \ref{fig3}. If $\mathcal{L}$ is an invertible defect associated with a global symmetry $g$ of $\mathcal{Q}$, the new operator $\hat{\mathcal{LO}}$ is nothing else the transformed operator over $g$ in its proper representation: $\hat{\mathcal{LO}}=R(g) \cdot \hat{\mathcal{O}}$. 
    \begin{figure}[h!]
        \centering
        \includegraphics[width=0.6\linewidth]{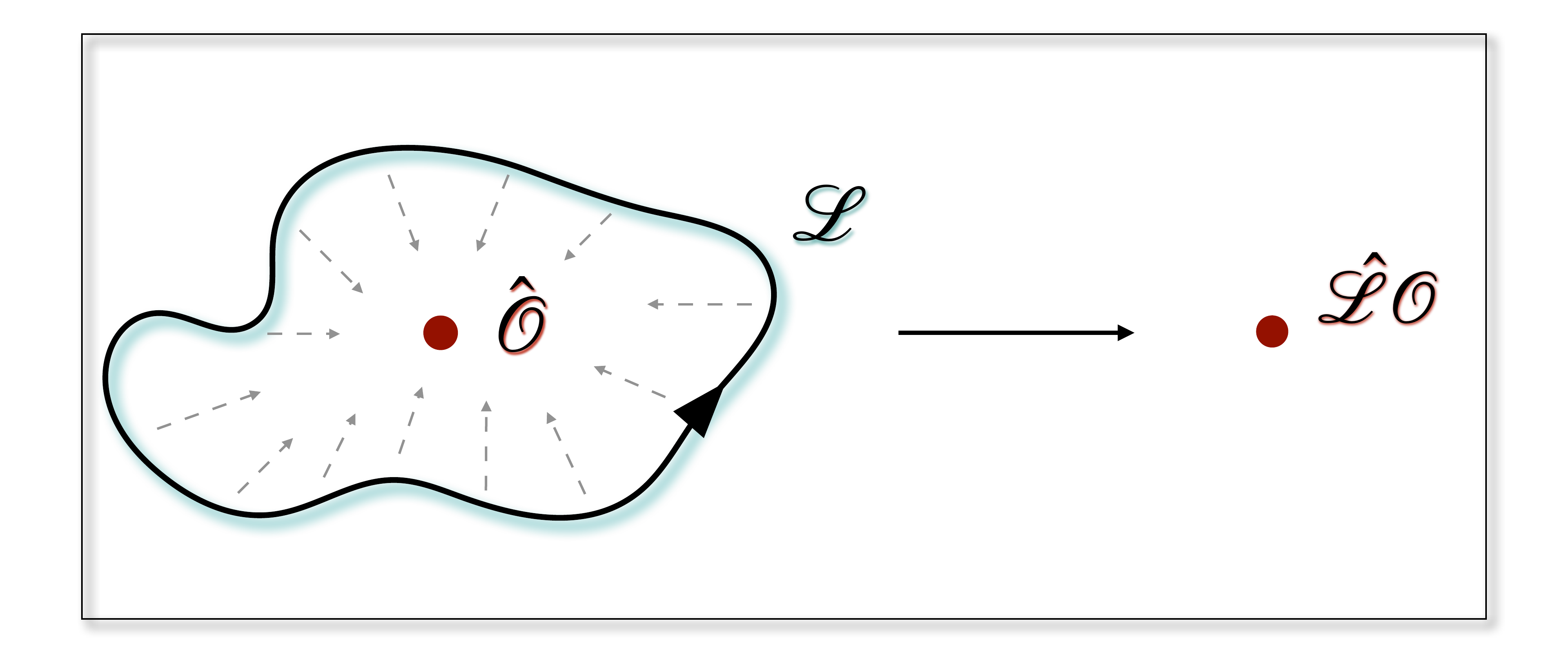}
        \caption{Action of the defect $\mathcal{L}$ on the local operator $\hat{\mathcal{O}}(z)$, producing the new operator $\hat{\mathcal{LO}}(z)$.}
        \label{fig3}
    \end{figure}
    
    \item Given the defects $\mathcal{L}_a$ and $\mathcal{L}_b$ in $S$, one can define a third defect through \textit{superposition}, such that $\widehat{\mathcal{L}_a + \mathcal{L}_b} = \hat{\mathcal{L}}_a + \hat{\mathcal{L}}_b $ and $\mathcal{H}_{\mathcal{L}_a + \mathcal{L}_b} = \mathcal{H}_{\mathcal{L}_a} \oplus \mathcal{H}_{\mathcal{L}_b}$. This operation is associative and commutative. 
    \item The set $S$ carries a second operation among its elements called \textit{fusion}. Given any two TDLs $\mathcal{L}_a, \mathcal{L}_b \in S$ one defines their fusion $\mathcal{L}_a \mathcal{L}_b$ by bringing their support lines into coincidence. The corresponding operator  is simply the product $\hat{\mathcal{L}_a \mathcal{L}_b} = \hat{\mathcal{L}}_a \hat{\mathcal{L}}_b$, while the new $\mathcal{L}_1 \mathcal{L}_2$-twisted sector is given by the tensor product $\mathcal{H}_{\mathcal{L}_a \mathcal{L}_b} = \mathcal{H}_{\mathcal{L}_a} \otimes \mathcal{H}_{\mathcal{L}_b}$. Fusion is associative, distributes with respect to the direct sum  and may or may not be commutative. \\ Equipped with the operations of superposition (direct sum) and fusion, the set $S$ acquires the structure of a (generally noncommutative) ring. 
    \item By inserting a TDL $\mathcal{L}$ into a correlation function $\langle \hat{\mathcal{L}} \, \hat{\mathcal{O}}_1 \, \hat{\mathcal{O}}_2 \, ... \rangle $ of the theory, one can analyze how the correlator behaves as $\mathcal{L}$ is moved across the insertion points of a local operator $\mathcal{O}_i$. If all correlators  remain unchanged under this operation, one say that the defect $\mathcal{L}$ and the local operator $\mathcal{O}_i$ are \textit{transparent} one to the other.  In a CFT, if moving $\mathcal{L}$  across a subset of local operators, corresponding to the holomorphic fields $\left\lbrace  \mathcal{O}_1(z), \mathcal{O}_2(z) , ... \right\rbrace$, leaves all correlation functions invariant, then the TDL $\mathcal{L}$ is said to preserve the corresponding chiral algebra $\mathcal{A}$ generated by these fields. An analogous statement holds for the anti-chiral algebra $\bar{\mathcal{A}}$ associated with anti-holomorphic fields. The full preserved algebra $\mathcal{A} \times \bar{\mathcal{A}}$ must necessarily include the Virasoro algebra $Vir_c \times Vir_{\bar{c}}$, generated by the stress-energy tensors $T(z)$ and $\bar{T}(\bar{z})$, and it is typically non-rational.    
    \item From a mathematical perspective, topological defects are expected to be described as objects in a fusion or, more generally, a tensor category. The morphisms inside this category are topological junction operators in the space $ Hom(\mathcal{L}_a, \mathcal{L}_b)$ connecting an incoming line $\mathcal{L}_a$ with an outgoing line $\mathcal{L}_b$. The word 'topological' means that the junction can be moved without changing any correlation function, as long as the deformation does not cross the support of another operator. Every element $u \in Hom(\mathcal{L}_a, \mathcal{L}_b)$ corresponds to a linear map $u: \mathcal{H}_{\mathcal{L}_a} \mapsto \mathcal{H}_{\mathcal{L}_b}$ among the corresponding twisted spaces. For each topological defect, there is always the trivial junction between $\mathcal{L}$ and itself, then $dim_{\mathbb{C}} Hom(\mathcal{L}, \mathcal{L}) \geq 1$. When this disequality is saturated, the defect $\mathcal{L}$ is said \textit{simple}. Beyond this, one can consider topological three-junction operators in the space $Hom(\mathcal{L}_a \otimes \mathcal{L}_b, \mathcal{L}_c)$, or more general $k$-way topological junctions.  
\end{itemize}
If we focus on CFTs, one of their defining features is the state–operator correspondence, which identifies each local operator $\hat{\mathcal{O}}_{\psi}(z)$ on the complex plane with a state $\vert \psi \rangle$ in the Hilbert space on the spatial circle $S^1$ of the cylinder.  This correspondence arises because the punctured plane and the cylinder are related by a conformal transformation $\mathbb{C}^{\ast}\simeq \mathbb{C} \setminus \left\lbrace 0 \right\rbrace \cong S^1 \times \mathbb{R}$. This property has profound implications, even for the interpretation of TDLs and their action on the states.  \\
Suppose to act with a topological defect $\mathcal{L}$ on a local operator $\mathcal{O}_{\psi}$ in the complex plane by enclosing it with an oriented closed line $\gamma$ supporting $\mathcal{L}$ and then shrinking $\gamma$ onto $\hat{\mathcal{O}}_{\psi}$. As previously discussed, this produces a new local operator $\widehat{\mathcal{L O}_{\psi}}$. Mapping  this configuration to the cylinder, as depicted in Figure \ref{fig4}, the defect $\mathcal{L}(\gamma)$ encircling $\hat{\mathcal{O}}_{\psi} (z)$ in the plane corresponds to $\mathcal{L}$ wrapping a spacelike $S^1$ on the cylinder. The retraction of $\gamma$ in the plane translates to moving $\mathcal{L}$  toward the asymptotic Hilbert space on the cylinder. Since the shrinking operation in the plane yields a new local operator, the corresponding effect on the cylinder is a linear transformation of the state $\vert \psi \rangle$, associated with the initial operator $\hat{\mathcal{O}}_{\psi}$, into a new state $\vert \hat{\mathcal{L}} \psi \rangle$, associated with the new operator $\widehat{\mathcal{L O}_{\psi}}$, on the same Hilbert space. In other words, each topological defect $\mathcal{L}$ defines a linear operator
\begin{equation}
    \hat{\mathcal{L}} \, \, : \, \, \mathcal{H} \, \, \, \longmapsto \, \, \, \mathcal{H}
\end{equation}
acting on the Hilbert space. Since $\mathcal{L}$ implements a generalized symmetry, the vacuum $\vert 0 \rangle$ must be an eigenvector of $\hat{\mathcal{L}}$. Its eigenvalue, $\langle \mathcal{L} \rangle$, is known as the \textit{quantum dimension} of the defect. 
\begin{figure}[h!]
    \centering
    \includegraphics[width=0.8\linewidth]{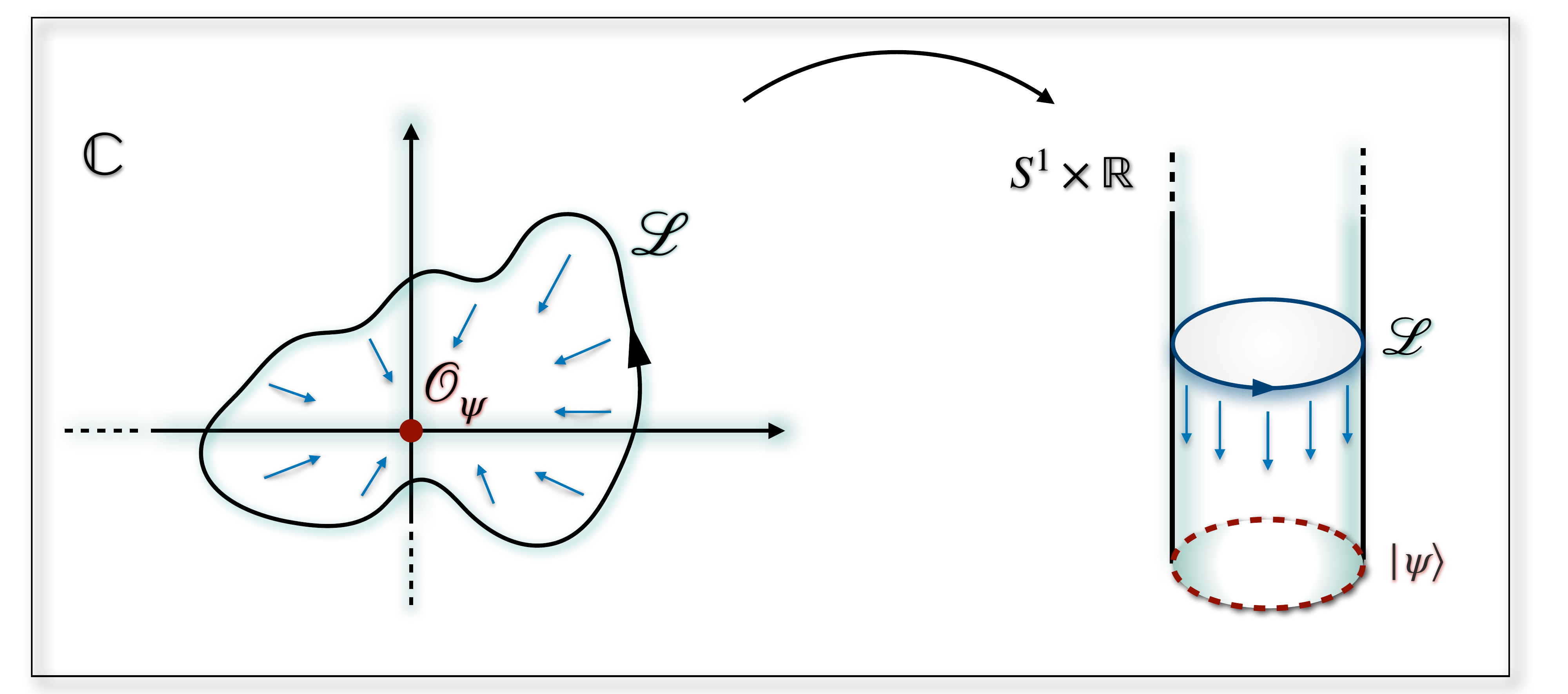}
    \caption{Conformal map sending the defect $\mathcal{L}$ encircling the local operator $\mathcal{O}_{\psi}$ on the complex plane to the same defect wrapped around the spatial $S^1$ on the cylinder.}
    \label{fig4}
\end{figure}

A second key implication arises when $\mathcal{L}$  is inserted along the timelike direction of the cylinder. This modifies the asymptotic Hilbert space $\mathcal{H}$ into a new space $\mathcal{H}_{\mathcal{L}}$, the $\mathcal{L}$-twisted space, whose states are twisted with respect to the action of $\hat{\mathcal{L}}$. Mapping back to the complex plane, each state  $\vert \psi \rangle \in \mathcal{H}_{\mathcal{L}}$ corresponds to a defect operator $\hat{\mathcal{O}}_{\psi}$ attached to an outgoing line supporting the operator $\hat{\mathcal{L}}$, see Figure \ref{fig5}. The space $\mathcal{H}_{\mathcal{L}}$ carries a representation of the preserved algebra $\mathcal{A} \times \bar{\mathcal{A}}$ preserved by the defect. If this representation is irreducible, then $\mathcal{L}$ is a simple defect. 
\begin{figure}[h!]
    \centering
    \includegraphics[width=0.7\linewidth]{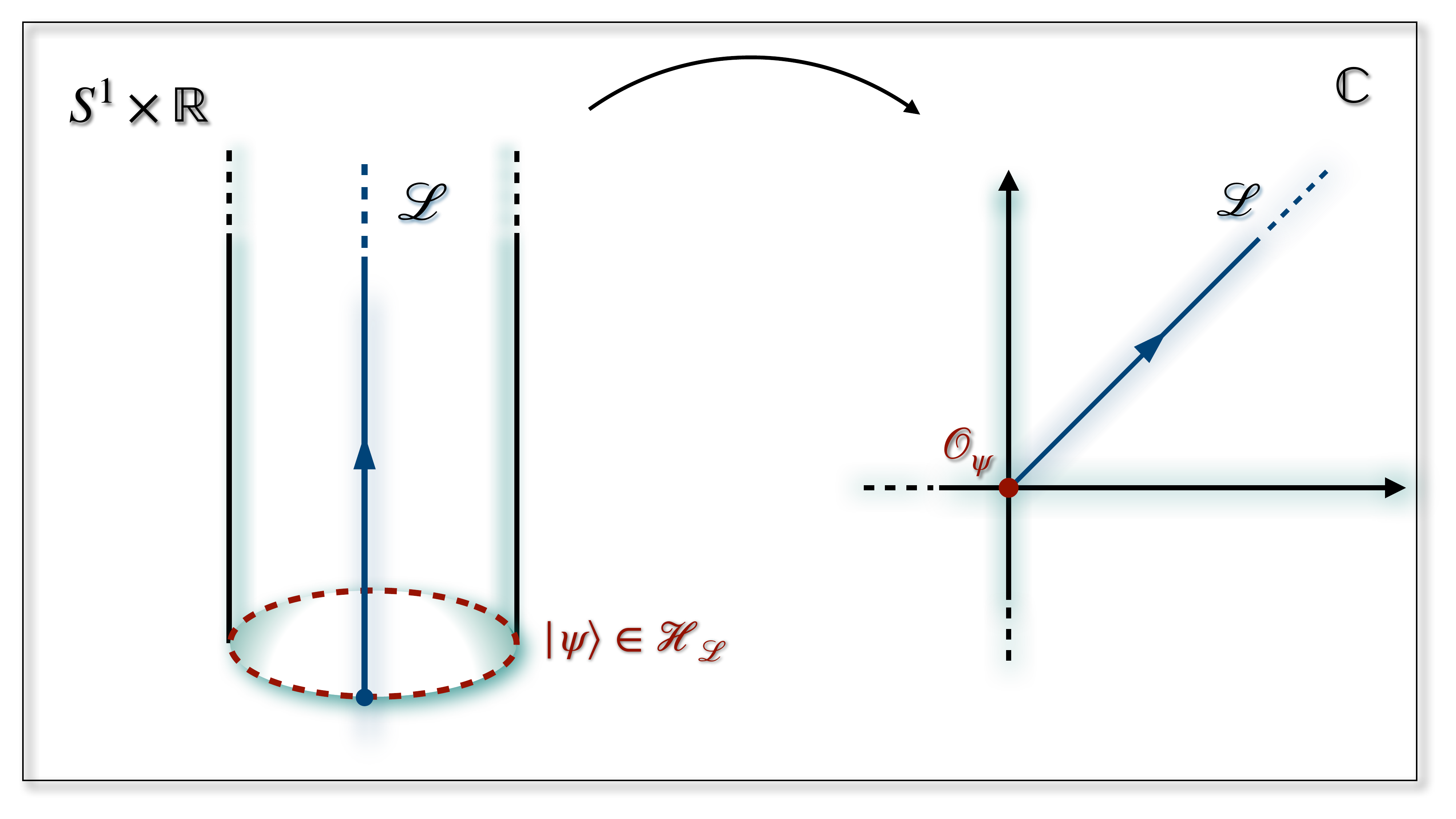}
    \caption{Conformal map sending the twisted state  $\vert \psi \rangle \in \mathcal{H}_{\mathcal{L}}$ on the cylinder to the corresponding defect operator $\hat{\mathcal{O}}_{\psi}$ on the complex plane.}
    \label{fig5}
\end{figure}

\subsection{State of the art}
The conceptual generalization of symmetries through categorical descriptions of topological defects has motivated extensive research into their properties and classification across diverse physical domains. Significant advances span quantum field theories, string theory \cite{Cordova:2023qei, Angius:2024evd, Apruzzi:2021nmk, Apruzzi:2022rei, Apruzzi:2023uma, Apruzzi:2024ark, Bhardwaj:2024qiv, Kaidi:2024wio, Heckman:2024obe, Bhardwaj:2025jtf, Bhardwaj:2025piv}, condensed matter systems \cite{Zhang:2023wlu, Cordova:2024goh, Cordova:2024mqg}, and lattice models \cite{Bhardwaj:2024kvy, Inamura:2025cum}. Despite substantial progress, a complete classification of these generalized symmetries remains elusive, even in relatively simple settings like conformal field theories (CFTs), where foundational results exist.

In two-dimensional CFTs, systematic classifications are currently confined to specialized cases:
\begin{itemize}
    \item \textit{Invertible defects,} which correspond to standard or higher form global symmetries of the theory.
    \item \textit{Verlinde lines} in Rational Conformal Field Theories (RCFTs) \cite{Verlinde:1988sn}, which are in one-to-one correspondence with the irreducible representations of the (anti-)chiral algebra and which fusion ring is equivalent to the fusion ring generated by the tensor product of that representations.
    \item \textit{Duality defects} \cite{Tambara:1998,Bhardwaj:2017xup}, which occur when a theory $\mathcal{C}$ is self-orbifold with respect to one proper discrete subgroup $H$ of symmetry, namely $\mathcal{C}/H \cong \mathcal{C}$. This means that there is an isomorphism between the space $\mathcal{H}_{\mathcal{C}}$ and $\mathcal{H}_{\mathcal{C}/H}$ of local operators of the two theories and their correlation function. The isomorphism is implemented by the presence of a duality defect in the theory satisfying the following fusion rules
    \begin{equation}
    \mathcal{N}^{\ast} \mathcal{N} = \sum_{h \in H} \mathcal{L}_h \, \, , \, \quad \, \mathcal{L}_h \mathcal{N} = \mathcal{N} \mathcal{L}_h = \mathcal{N}.
      \label{fusion_rules:TY}  
    \end{equation}
    If $\mathcal{N}$ is unoriented, i.e. $\mathcal{N}^{\ast} = \mathcal{N}$, then the category with objects the set $\left\lbrace \mathcal{I}, \mathcal{N}, \mathcal{L}_h \right\rbrace$, for $h \in H$, equipped with the fusion rules \eqref{fusion_rules:TY}  and the invertible-defect fusion law induced by the group multiplication of $H$, forms a so called Tambara-Yamagami (TY) category.
\end{itemize}
Systematic approaches for analyzing general topological defects in non-rational CFTs remain largely undeveloped. However, K3 NLSMs offer a promising starting point for such investigations. Although these models are non-rational, they possess an exceptionally large amount of worldsheet supersymmetry, which imposes powerful constraints on their structure. This makes K3 models a particularly tractable class for making significant progress in understanding topological defects in non-rational settings.

\section{General results about TDLs in K3 models} \label{Section:4}
In this section, we generalize the method described in subsection \ref{sec:K3NLSM} to characterize the symmetries of $K3$ NLSMs. In particular, our purpose is to characterize the category $\cTop_\Pi$ of topological defects of $\calC_\Pi$ that preserve
the $\mathcal{N}=(4,4)$  superconformal algebra and act trivially on the $4$-dimensional space $\Pi$ spanned by the spectral flow generators. Under these assumptions, TDLs preserve spacetime supersymmetry so that they map 1/2 BPS D-branes into 1/2 BPS D-branes, and RR charge vectors to RR charge vectors. One can therefore argue that there is a mapping
\begin{align}\label{homtoend}
    \cTop_\Pi &\to {\rm End}(\Gamma^{4,20})\\
    \CL&\mapsto \nL\notag\,
\end{align}
from a TDL $\CL$ to a $\mathbb{Z}$-linear function $\nL: \Gamma^{4,20} \to \Gamma^{4,20}$, preserving the fusion ring structure of $\cTop_\Pi$. We denote by the same symbol $\nL$ the extension by $\mathbb{R}$-linearity of this lattice endomorphism to the real space of R-R ground fields $V\cong \Gamma^{4,20}\otimes \RR$. Since $\CL$ commutes with the $\mathcal{N}=(4,4)$ superalgebra, it cannot cannot mix RR ground states in different representations, and its trivial action on the spectral flow generartors must be the same as the one on the vacuum, i.e. by multiplication by the quantum dimension $\langle \CL\rangle \geq 1$. It therefore follows that
$\nL:V\to V$ is block-diagonal with respect to the orthogonal decomposition $V=\Pi\oplus \Pi^\perp$, i.e. $\nL(\Pi)\subseteq \Pi$ and $\nL(\Pi^\perp)\subseteq\Pi^\perp$. Furthermore, the restriction $\nL_{\rvert \Pi}$ is proportional to the identity $\nL_{\rvert\Pi}=\langle \CL\rangle{\rm id}_\Pi$, where $\langle \CL\rangle\geq 1$ is the quantum dimension of the defect. Choosing a suitable orthonormal basis of $V$, compatible with the splitting $V\cong \Pi\oplus\Pi^\perp$, we can characterize $\nL$ as a matrix in the real vector space (having the structure of a semiring)
\be\label{blkspaceplus}
B^{4,20}_+(\RR):=\left\{\left(\begin{smallmatrix}
    d\cdot \mathbf{1}_{4\times 4} & 0\\
    0 & b_{20\times 20}
\end{smallmatrix}\right)\mid d\ge 1, b_{20,20}\in Mat_{20\times 20}(\RR)\right\}\subset B^{4,20}(\RR)\ .\ee
The semiring homomorphism
\begin{align}\label{homtoB}
    \cTop_\Pi &\to B^{4,20}_{\Pi,+}(\ZZ)\subset {\rm End}(\Gamma^{4,20})\\
    \CL&\mapsto \nL\,,\notag
\end{align}
where $B^{4,20}_{\Pi,+}(\ZZ):={\rm End}(\Gamma^{4,20})\cap B^{4,20}_+(\RR)$, can be understood as an extension of the group  isomorphism $\rho:G_\Pi\to {\rm Stab}(\Pi)$. However, whereas  a TDL $\mathcal{L}$ is invertible if and only if its  quantum dimension is $\langle \CL\rangle = 1$, it is evident the condition $d=1$ is also consistent with non-invertible $L\in B^{4,20}_{\Pi,+}(\ZZ) $, implying the map \eqref{homtoB} cannot be surjiective. Having defined $\mathsf{K}_\Pi$ as the subcategory of topological defects $\CL$ of the model $\calC_\Pi$ preserving $\CN=(4,4)$ and spectral flow, and such that
\be \rho(\mathsf{K}_\Pi)\subseteq\{\nL=d\cdot {\rm id}_V,\ d\ge 1\}\subset B^{4,20}_{\Pi,+}(\ZZ)\ ,
\ee
one can  find the following result, which generalizes to the category $\cTop_\Pi$ the fact that the identity element in ${\rm Stab}(\Pi)$ corresponds the trivial identity symmetry in $G_\Pi$.
\begin{claim}\label{th:onlyid}
    For all K3 models $\calC_\Pi$ the category $\mathsf{K}_\Pi$  of defects preserving $\CN=(4,4)$ and spectral flow, and acting by multiplication by some $d\in \RR$ on the space of RR ground fields $V$ is generated by the trivial defect
    \be \mathsf{K}_\Pi=\{d\CI,\ d\in \NN\}\ .
    \ee
\end{claim}
In fact, first we notice that the real number $d$ must be the quantum dimension $d=\langle \CL\rangle$, and that therefore all R-R ground states are transparent to $\CL$. Then $\CL$ is transparent to any marginal deformation of the NLSM $\calC_\Pi$, implying that $\mathsf{K}_\Pi$ is the same for any point in the $K3$ moduli space, which is simply connected. In particular, in section \ref{sec:Example} we  discuss the $K3$ model with maximal symmetry $\ZZ_2^8:M_{20}$, $\mathcal{C}_{GTVW}$, for which one can argue that \emph{all} operators in $\CH$ can be obtained by OPE of R-R operators in $V$ and their $\CN=(4,4)$ descendants. In this model, the bosonic parts of the holomorphic and anti-holomorphic chiral algebras are isomorphic to $\CA:=(\widehat{su}(2)_1)^{6}$ (six copies of the $su(2)$ current algebra at level $1$), and the $24$ (real) Ramond-Ramond states with conformal weight $h=\bar h=\frac{1}{4}$ generating the space $V$ are the ground states in the modules \ref{eq:RRreps} described in section \eqref{sec:Example}, so that they are  naturally organized in tetrads. We can identify the $\widehat{su}(2)_1$ subalgebra of the (anti-)holomorphic $\CN=4$ with the first factor in the (anti-)chiral algebra $(\widehat{su}(2)_1)^{6}$. In fact, by taking the OPE of the four operators the $i$-th tetrad, $i=1,\ldots, 6$, one can obtain the currents in the holomorphic and anti-holomorphic $i$-th $\widehat{su}(2)_1$ factor. Thus, in this way we can generate the full chiral and antichiral algebra $\CA\times\bar\CA$. Furthermore, the fusion rules between $\CA$ modules imply that the OPE of the $\CA\times\bar\CA$ primary fields in the six modules \eqref{eq:RRreps}, together with the ones in the $[1,1,1,1,1,1;0,0,0,0,0,0]$ and $[0,0,0,0,0,0; 1,1,1,1,1,1]$ modules, where the supercurrents live, generate every other $\CA\times\bar\CA$ primary operator in the spectrum. In this way, 
one can obtain \emph{all} operators in $\CH$ by OPE of R-R operators in $V$ and their $\CN=(4,4)$ descendants. A defect $\CL$ that is transparent to the $\CN=(4,4)$ algebra and to all R-R operators in $V$ is therefore transparent to all operators in  $\CH$. 

This result is however not equivalent to the injectivity of the map \eqref{homtoB}, since two defects can be mapped to the same $\nL\in B^{4,20}_{\Pi,+}(\ZZ)$ without being obtainable from each other by fusion with a defect in $\mathsf{K}$.

However the map \eqref{homtoB} can still be used to further characterize TDLs. In particular, if 
 $\CL\in\cTop_{\calC_\Pi}$, andthere is some $\Psi\in \Gamma^{4,20}$, $\Psi\neq 0$, such that $\nL(\Psi)=\langle \CL\rangle\Psi$, where $\langle\CL\rangle$ is the quantum dimension of $\CL$, then  $\langle \CL\rangle$ is integral since $\Psi\in \Gamma^{4,20}$ can be taken primitive (i.e., not an integral multiple of a shorter vector in the lattice) with no loss of generality and $\langle\CL\rangle\Psi\in \Gamma^{4,20}$ must necessarily be an integral multiple of $\Psi$.
Indeed, a stronger constraint can be derived observing that $p(\nL)=0$ where $p(x)$, the characteristic polynomial, is a monic polynomial of degree $24$ with integral coefficients. In particular, $\langle \CL\rangle$, which is an eigenvalue of $\nL$, is also a root of the same monic polynomial, i.e. it is an algebraic integer. If $r(x)$ is the minimal polynomial for $d=\langle \CL\rangle$, i.e. the least degree monic polynomial with integral coefficient having $d$ as a root, $r(x)^4$ divides $p(x)$ because $d$ has multiplicity at least $4$. It follows that $r(x)$ must have degree at most $6$, and the following facts can be stated.
\begin{claim}\label{th:qdim}
    The quantum dimension $\langle \CL\rangle$ of a defect $\CL\in \cTop_\Pi$ is an algebraic integer of degree at most $6$. Furthermore, if $\Pi\cap \Gamma^{4,20}\neq 0$, then $\langle \CL\rangle$ is integral for all $\CL\in \cTop_{\Pi}$.
\end{claim}
We can notice the condition $\Pi\cap \Gamma^{4,20}\neq 0$ can be physically interpreted.  From the viewpoint of type IIA superstring, a primitive vector $v\in \Gamma^{4,20}$ with $v^2>0$ represents the charge of a BPS D0-D2-D4-brane configuration. The mass of such a BPS configuration depends on the moduli, and is proportional to $v_\Pi^2$, where $v_\Pi$ and $v_\perp$ are the orthogonal projections of $v$ along $\Pi$ and $\Pi^\perp$, so that $v^2=v_\Pi^2-v_\perp^2$. An attractor point in the moduli space for the BPS state with charge $v$ is a point where the BPS mass $v_\Pi^2=v^2+v_\perp^2$ is minimized, and  this happens if and only if $v\in \Pi$ \cite{Andrianopoli:1998qg,Dijkgraaf:1998gf,Ferrara:1995ih,Moore:1998pn}. Thus, the points in the moduli space where $\Pi \cap  \Gamma^{4,20}\neq 0$ are exactly the attractor points for some BPS brane configurations. Claim \ref{th:qdim} then implies that whenever the K3 model $\calC$ is `attractive', all topological defects $\CL\in \cTop_\calC$ have integral quantum dimension.

An interesting consequence then follows. Given $\Psi_1,\ldots,\Psi_{24}$ generators of the lattice $\Gamma^{4,20}\subset V$, and a defect $\CL\in \cTop_\Pi$ with quantum dimension $d:=\langle \CL\rangle$, for a generic model $\calC_\Pi$, we expect the scalar products $\langle \psi|\Psi_1\rangle,\ldots,\langle \psi|\Psi_{24}\rangle$, with some spectral flow generator $\psi\in \Pi$ to be linearly independent  over $\QQ[d]$, the extension of the field $\QQ$ by the algebraic integer $d$.
Using
\be d\langle\psi|\Psi_i\rangle= \langle\psi|\nL|\Psi_i\rangle=\sum_{j=1}^{24} \nL_{ij} \langle\psi|\Psi_j\rangle\ ,
\ee we get
\be \sum_j (d\delta_{ij}-\nL_{ij})\langle\psi|\Psi_j\rangle=0\qquad \forall i=1,\ldots,24\ ,
\ee
because $d\delta_{ij}-\nL_{ij}\in \QQ[d]$, we therefore get \be d\delta_{ij}-\nL_{ij}=0\qquad \forall i,j=1,\ldots,24\ .
\ee
This means that $\nL$ is $d$ times the identity, and by proposition \ref{th:intdim} $d$ is integral. By Claim \ref{th:onlyid}, this means that $\CL=d\CI$.
\begin{claim}\label{th:generic}
For a generic K3 sigma model $\calC_\Pi$, 
the only topological defects in $\cTop_\Pi$ are integral multiples of the identity. 
\end{claim}
We are therefore driven to the conclusion that for any nontrivial TDL $\CL\in \cTop_\Pi$, there is a deformation of the $K3$ NLSM $\calC_\Pi$ lifting it. This happens in particular for the $K3$ models having a nontrivial symmetry group ${\rm Stab}(\Pi)$. The result is not even in contradiction with the subset of the moduli space where $\cTop$ is nontrivial being dense in $\CM_{K3}$.

A particularly interesting family of $K3$ models having a nontrivial $\cTop$ including non-invertible defects are the ones that can be described as orbifolds of supersymmetric sigma models on $T^4$ by some finite group of symmetries. For a generic $T^4$ model $\CT$, the group $G_\CT$ of symmetries preserving the small $\CN=(4,4)$ algebra and the spectral flow generators is 
\be G_\CT\cong U(1)^8\rtimes \ZZ_2\ .
\ee 
$U(1)^8\cong U(1)^4\times U(1)^4$ is the product of the $U(1)^4$ group of translations along the four direction of the torus $T^4$, times the $U(1)^4$ group of translations along the T-dual torus. The action of 
\be W_\theta\ ,\qquad\qquad \theta\in (\Gamma^{4,4}\otimes\RR)/\Gamma^{4,4}\cong (\RR/\ZZ)^8 \ ,\ee
on primary states $|\lambda\rangle$ labeled by 
$\lambda\in \Gamma^{4,4}$ (the Narain lattice)
is given by
\be \hat W_\theta|\lambda\rangle= e^{2\pi i (\theta,\lambda)}|\lambda\rangle\ ,.
\ee 
The $\ZZ_2$ symmetry $\calR$ is the coordinate reflection
\be\label{refl} \partial X^k\to -\partial X^k\ ,\qquad \bar\partial X^k\to -\bar\partial X^k\qquad \psi^k\to-\psi^k\ ,\qquad\tilde\psi^k\to-\tilde\psi^k\ .
\ee The fusion of $\calR$ with $W_\theta$ is
\be \calR W_\theta=W_{-\theta}\calR\ .
\ee
A torus orbifold $\calC=\CT/\ZZ_2$  by the $\ZZ_2$ symmetry \eqref{refl} always admits an invertible defect $Q$ (the quantum symmetry) acting by $-1$ on the twisted sector and trivially on the untwisted one. Besides $Q$, the subcategory of invertible symmetries is  generated by the simple defects that are induced by the simple defects $W_\theta$ of the torus model $\CT$ that commute with $\calR$, i.e such that $\theta\equiv -\theta\mod \Gamma^{4,4}$. So, for each defect $W_{\frac{\lambda}{2}}$, $\lambda\in \Gamma^{4,4}/2\Gamma^{4,4}$ of $\CT$, there are two  associated defects
$\eta_\frac{\lambda}{2},\eta'_{\frac{\lambda}{2}}\equiv Q\eta_\frac{\lambda}{2}$ of $\calC$, with fusion rules
\be Q^2=I\ ,\qquad  \eta_{\frac{\lambda}{2}} Q=Q\eta_{\frac{\lambda}{2}} \ ,\qquad (\eta_{\frac{\lambda}{2}})^2=Q^{(\lambda,\lambda)/2}
\ee  which imply
\be \eta_{\frac{n_i}{2}}^2=I=\eta_{\frac{w_i}{2}}^2\ ,\qquad \eta_{\frac{\lambda}{2}}\eta_{\frac{\mu}{2}}=Q^{(\lambda,\mu)}\eta_{\frac{\mu}{2}}\eta_{\frac{\lambda}{2}}\,.
\ee
Whereas the defects $W_{\frac{\lambda}{2}}$, $\lambda\in \Gamma^{4,4}$, generate an abelian group of symmetries $\ZZ_2^8\subset U(1)^8$ of the torus model $\CT$, the group generated by the $\eta_{\lambda/2}$ is a non-abelian extension of $\ZZ_2^8$ by a central $\ZZ_2$ generated by the quantum symmetry $Q$,  corresponding to an extraspecial group $2^{1+8}$ \cite{Atlas}
\be 1\longrightarrow \langle Q\rangle\cong\ZZ_2 \longrightarrow 2^{1+8}\longrightarrow \ZZ_2^8\longrightarrow 1\ .
\ee
Torus orbifolds $\CT/\ZZ_2$ also contain a continuum of defects $T_\theta$, parametrized by $\theta \in ((\RR/ \ZZ)^8)/{\pm 1}$, that preserve the $\CN=(4,4)$ algebra and the spectral flow generators, and that are induced by the $\calR$-invariant superposition $W_\theta+W_{-\theta}$ of topological defects  of the torus model $\CT$. The defects $T_\theta$ have dimension $2$ and satisfy the fusion rules
$$ T_\theta T_{\theta'}=T_{\theta+\theta'}+T_{\theta-\theta'}\ ,
$$ 
They are simple except for $\theta=\frac{\lambda}{2} \in \frac{1}{2}\Gamma^{4,4}/\Gamma^{4,4}$, when we have
\be T_\frac{\lambda}{2}=\eta_\frac{\lambda}{2}+Q\eta_\frac{\lambda}{2}\ .
\ee 
Since $T_0=\CI+Q$, one finds 
\be\label{Tthetasquare} (T_\theta)^2=T_0+T_{2\theta}=\CI+Q+T_{2\theta}\ ,
\ee that implies that $T_\theta$ is unoriented, $(T_\theta)^\dual=T_{\theta}$. The fusion with the invertible defects is
\be QT_\theta=T_\theta Q=T_\theta\ ,\qquad \eta_{\frac{\lambda}{2}}T_\theta=T_\theta\eta_{\frac{\lambda}{2}}=T_{\theta+\frac{\lambda}{2}}\ ,
\ee where in the last identity, one uses $\frac{\lambda}{2}\equiv -\frac{\lambda}{2} \mod \Gamma^{4,4}$, so that $T_{\theta+\frac{\lambda}{2}}=T_{\theta-\frac{\lambda}{2}}$.

According to \cite{Thorngren:2021yso}, the operators $\hat T_{\theta}$ associated with these defects act on the untwisted sector as the operators $\hat W_\theta+\hat W_{-\theta}$ in the original theory, whereas they annihilate the twisted sector. On the RR ground states, all $\hat T_\theta$ act by twice the identity on the states in the untwisted sector, and annihilate all the states in the twisted sector.

If $2\theta\in (\Gamma^{4,4}\otimes\RR)/\Gamma^{4,4}$ is a  $\calR$-fixed point, i.e. if $2\theta\equiv -2\theta\mod \Gamma^{4,4}$, then $\theta=\frac{\lambda}{4}$ for some $\lambda\in \Gamma^{4,4}$, and the fusion product \eqref{Tthetasquare} becomes
\be\label{lambdafour} (T_\frac{\lambda}{4})^2=\CI+Q+\eta_{\frac{\lambda}{2}}+Q\eta_{\frac{\lambda}{2}}\ .
\ee The right-hand side is just the sum over all invertible defects in the order $4$ group generated by $Q$ and $\eta_{\frac{\lambda}{2}}$ (this is either $\ZZ_2\times \ZZ_2$ or $\ZZ_4$, depending on whether $(\eta_\frac{\lambda}{2})^2$ equals $\CI$ or $Q$, i.e. if $(\lambda,\lambda)$ equals $0$ or $2$ mod $4$). Therefore, $T_{\frac{\lambda}{4}}$ is a duality defect, providing an equivalence of our theory to the orbifold by this subgroup. 

This analysis can be generalized to orbifolds $\CT/\ZZ_N$ of a torus model $\CT$ by a cyclic group $\langle g\rangle\cong\ZZ_N$ \cite{Angius:2024evd}.

As in claim \ref{th:generic} we have argued that a generic $K3$ model has a trivial $\cTop_\calC$, we are lead to the following conjecture characterizing $K3$ model in terms of the presence of a continuous family of topological defects $\CL_\lambda\in \cTop_\calC$, that preserve the $\CN=(4,4)$ algebra and spectral flow.
\begin{conjecture}\label{th:onlytori} Let $\calC$ be a K3 model that admits a continuous family of topological defects $\CL_\lambda\in \cTop_\calC$, preserving the $\CN=(4,4)$ superconformal algebra and spectral flow, and simple for all $\lambda$ except possibly a zero measure set. Then, $\calC$ is the (generalised) orbifold of a torus model $\CT$.
\end{conjecture}
As argued in \cite{Thorngren:2021yso}, a continuum of operators $\CL_\lambda$ is related to the presence of a conserved current $j\in \CH_{\CL_\lambda\CL_\lambda^\dual}$:
\be j:=J(z)dz+\tilde J(\bar z)d\bar z\ ,
\ee where $\bar \partial J(z)=0=\partial \tilde J(\bar z)$. $j\in \CH_{\CL_\lambda\CL_\lambda^\dual}\cong \CH_{\CL_\lambda}\otimes\CH^*_{\CL_\lambda}$ is interpreted as a linear operator from $\CH_{\CL_\lambda}$ to itself.
The NS-NS sector of $\CH_{\CL_\lambda\CL_\lambda^\dual}$ contains two $\CN=(4,4)$ BPS representations $(\frac{1}{2},\frac{1}{2};0,0)$ and $(0,0;\frac{1}{2},\frac{1}{2})$, containing the currents $J(z)$ and $\tilde J(\bar z)$.
Taking then $\CL_{orb}\subseteq \CL_\lambda\CL_\lambda^\dual$ to be the smallest unoriented ($\CL_{orb}^\dual=\CL_{orb}$) defect containing $I$, with $\CL_{orb}\subset (\CL_{orb})^2$, and such that $\CH_{\CL_{orb}}$ contains the four holomorphic and antiholomorphic spin $1/2$ fields and is closed under OPE, $\CH_{\CL_{orb}}$ contains the chiral algebra of a generic $T^4$ model.
If $\calC/A$ is well-defined, with  $A:=\CL_{orb}$, it is a generalised orbifold of the torus model $\CT$ and by reversibility of the orbifold procedure, one can conclude that $\calC$ is a generalised orbifold of the torus model $\CT$.

In the case of the torus orbifolds $\CT/\ZZ_2$, $\CL_{orb}$ was given by $I+Q$, and indeed the orbifold of $\calC$ by $\CL_{orb}$ gives back $\CT$.

\subsection{The connection with $V^{f \natural}$}

Motivated by the formally analogy between the properties of topological defects in $V^{f \natural}$, stated in \cite{Angius:2025zlm}, and the ones in $K3$ models \cite{Angius:2024evd}, reviewed above, the authors of \cite{Angius:2025zlm, Angius:2025xxx} propose a generalization of the relationship between the two theories, extending it from the level of ordinary symmetries to that of their associated tensor categories. In particular, for any defect $\mathcal{L} \in Top_{\Pi^{\natural}}$, where $Top_{\Pi^{\natural}}$ is the subcategory of $Top$ of objects in $V^{f \natural}$ that fix a subalgebra $\hat{so}(4)_1 \subset \hat{so} (24)_1$ related to the four-dimensional space $\Pi^{\natural} \subset \mathbf{24}$, we can define a generalized $\mathcal{L}$-twined partition function:
\begin{equation}
    \phi^g \left( V^{f \natural}, \tau, z \right) = Tr_{V^{f \natural}_{tw}} \left( \hat{\mathcal{L}} \,(-1)^F q^{L_0 - 1/2} y^{J^3_0}\right)
    \label{Vfnat:L_twined_graded_part_func}
\end{equation}
taking trace of the action of $\mathcal{L}$ on the states of the twisted module $V^{f \natural}_{tw}$. Similarly, given the K3 model $\mathcal{C}$, for each topological defect $\mathcal{L} \in Top^{K3}_{\mathcal{C}}$, we can define the $\mathcal{L}$-twined elliptic genus:
\begin{equation}
    \phi^g \left( \mathcal{C}, \tau, z \right) = Tr_{RR} \left(\hat{\mathcal{L}} \, (-1)^{F + \bar{F}} q^{L_0-1/4} \bar{q}^{\bar{L}_0 -1/4} y^{J^3_0}\right).
    \label{K3:g_twined_elliptic_genus}
\end{equation}
The expectation of an exact defect correspondence is formalized in the following conjecture (Conjecture 4 in \cite{Angius:2025zlm}).
\begin{conjecture}\label{conj:K3relation}
    Let $\Pi^{\natural} \subset \Lambda\otimes\RR\subset V^{f\natural}_{tw}$ be a subspace of dimension $4$ and $\cTop_{\Pi^{\natural}}$ the full subcategory of $\cTop$ such that
    \begin{equation}
        Obj \left( Top_{\Pi^{\natural}}\right) \, = \, \left\lbrace \mathcal{L} \in Obj (Top) \, \vert \, \hat{\mathcal{L}}\vert_{\Pi^{\natural}} = \langle \mathcal{L} \rangle \text{id}_{\Pi^{\natural}}\right\rbrace.
    \end{equation}
     Then, there exists a non-linear sigma model $\calC$ on K3 such that:\begin{enumerate}
        \item[\textit{(i)}] there in an equivalence $F:\cTop_{\Pi^{\natural}}\longrightarrow \cTop^{K3}_\calC$ of tensor categories between $\cTop_{\Pi^{\natural}}$ and the category $\cTop^{K3}_\calC$ of topological defects of $\calC$ preserving $\CN=(4,4)$ and spectral flow;
        \item[\textit{(ii)}] for every $\CL\in \cTop_{\Pi^{\natural}}$, the $\CL$-twining genus $\phi^\CL(V^{f\natural},\tau,z)$  computed in $V^{f\natural}$ (eq.\eqref{Vfnat:L_twined_graded_part_func}) coincides with the $F(\CL)$-twining genus in the K3 model $\calC$ (eq.\eqref{K3:g_twined_elliptic_genus}):
 \be        \phi^\CL(V^{f\natural},\tau,z)=\phi^{F(\CL)}(\calC,\tau,z).
        \ee 
\item[\textit{(iii)}] there is an isometry of Hilbert spaces $\varphi:V^{f\natural}_{tw}(1/2)\stackrel{\cong}{\longrightarrow} \Hh^{K3}_{\HRR,gr}$ such that for each $\CL\in \cTop_{\Pi^{\natural}}$
\be \varphi\circ \hat\CL_{\rvert V^{f\natural}_{tw}(1/2)} =\widehat{F(\CL)}_{\rvert \Hh^{K3}_{\HRR,gr}}\circ \varphi\ .
\ee
    \end{enumerate}
\end{conjecture}

\section{Example: the $K3$ $\mathcal{C}_{GTVW}$ model}
\label{sec:Example}
In this section, we apply the general framework descibed above to classify topological defects that commute with the  $\mathcal{N}=(4,4)$ superconformal algebra and spectral flow in the highly symmetric K3 model  $\mathcal{C}_{GTVW}$ \cite{Gaberdiel:2013psa}. This model has several equivalent realizations: as a $\mathbb{Z}_2$ orbifold of the $D_4$ torus theory under the reflection $\mathcal{R}$ in all four toroidal directions, or as an asymmetric $\mathbb{Z}_4$ orbifold of the same $D_4$ torus theory. \\
Equivalently, the model also admits a Rational Conformal Field Theory (RCFT) based on the chiral algebra $ \mathcal{A} \times \bar{\mathcal{A}} =\hat{su}(2)_{L,1}^6 \oplus \hat{su}(2)_{R,1}^6$. The algebra $\hat{su}(2)_1$ has exactly two irreducible representations: the vacuum representation with conformal weight $0$, and a non-trivial representation with conformal weight $1/4$. These representations fuse according to the group ring of $\mathbb{Z}_2$, which we represent as the additive group with elements $0,1$. Consequently, we can label each irreducible module $M_{\left[ a_1, ..., a_6\right]}$ of the rational algebra $\mathcal{A}$  by a six-component vector $\left[ a_1, ..., a_6 \right]$ in $\mathbb{Z}_2^6$. The NS-NS sector of the theory is given by the following irreducible module decomposition
\begin{equation}
    \bigoplus_{\left[ a_1, ..., a_6 ; b_1, ..., b_6\right] \in A_{NS-NS}} M_{[a_1,...,a_6]} \otimes \bar{M}_{\left[ b_1,...,b_6\right]},
    \label{GTVW:module_dec}
\end{equation}
where the set
\begin{equation}
   A_{NS-NS} = A_{(NS-NS)^+} \, \sqcup  \, A_{(NS-NS)^-} 
\end{equation}
contains a bosonic part
\begin{equation}
   A_{(NS-NS)^+} = \left\lbrace \left[a_1, ..., a_6; b_1, ..., b_6 \right] \in \mathbb{Z}_2^6 \times \mathbb{Z}_2^6 \, \vert \, a_i=b_i \, , \, \sum_i a_i \equiv 0 \, \text{mod} \, 2\right\rbrace 
\end{equation}
and a fermionic one
\begin{equation}
    A_{(NS-NS)^-} = \left\lbrace \left[a_1, ..., a_6; b_1, ..., b_6 \right] \in \mathbb{Z}_2^6 \times \mathbb{Z}_2^6 \, \vert \, a_i=b_i +1 \, , \, \sum_i a_i \equiv 0 \, \text{mod} \, 2\right\rbrace.
\end{equation}
Similarly, for the R-R sector we have the analogous decomposition \eqref{GTVW:module_dec} with respect to the set $A_{R-R}$ containing the bosonic subset
\begin{equation}
    A_{(R-R)^+} = \left\lbrace \left[a_1, ..., a_6; b_1, ..., b_6 \right] \in \mathbb{Z}_2^6 \times \mathbb{Z}_2^6 \, \vert \, a_i=b_i \, , \, \sum_i a_i \equiv 1 \, \text{mod} \, 2\right\rbrace 
\end{equation}
and the fermionic one
\begin{equation}
      A_{(R-R)^-} = \left\lbrace \left[a_1, ..., a_6; b_1, ..., b_6 \right] \in \mathbb{Z}_2^6 \times \mathbb{Z}_2^6 \, \vert \, a_i=b_i +1 \, , \, \sum_i a_i \equiv 1 \, \text{mod} \, 2\right\rbrace   .
\end{equation}
This model contains different copies of the $\mathcal{N}=(4,4)$ algebra, corresponding to different choices of $\hat{su}(2)_1$ subalgebras in both the holomorphic and anti-holomorphic sectors. Focusing on the first  $\hat{su}(2)_1$ in each sector, the $24$ R-R ground states organize into six irreducible representations:
\begin{equation}
\label{eq:RRreps}
    \begin{split}
        & \left[ 1,0,0,0,0,0;1,0,0,0,0,0 \right] \, \, , \, \, \left[ 0,1,0,0,0,0;0,1,0,0,0,0 \right] \, \, , \, \, \left[ 0,0,1,0,0,0;0,0,1,0,0,0 \right] \\
        & \left[ 0,0,0,1,0,0;0,0,0,1,0,0 \right] \, \, , \, \, \left[ 0,0,0,0,1,0;0,0,0,0,1,0 \right] \, \, , \, \, \left[ 0,0,0,0,0,1;0,0,0,0,0,1 \right] \\
    \end{split}
\end{equation}
where each representation contains exactly four ground states forming a tetrad. Notably, the first tetrad comprises the spectral flow generators spanning $\Pi \in \mathbb{R}^{4,20}$. The four holomorphic and anti-holomorphic supercurrents are suitable ground states chosen in the representations
\begin{equation}
   \left[ 1,1,1,1,1,1;0,0,0,0,0,0 \right] \, \, , \, \, \left[ 0,0,0,0,0,0;1,1,1,1,1,1 \right]  
\end{equation}
respectively.\\
The faithful symmetry group acting on the Hilbert space is given by
\begin{equation}
G=\left( ( SU(2)^6 \times SU(2)^6)/ \mathbb{Z}_2^5 \right) \rtimes S^6,
\end{equation}
where the two $SU(2)^6$ factors are generated by the zero modes of the chiral algebra currents, and $S^6$ is the diagonal permutation group acting simultaneously on the six $\hat{su}(2)^6_1$ factors in both the holomorphic and anti-holomorphic sectors. The subgroup preserving the super-Virasoro algebra and spectral flow is the split extension $G_{GTVW} \cong \mathbb{Z}_2^8: M_{20}$ of the Mathieu group $M_{20}$ by $\mathbb{Z}_2^8$. 
\subsection{Topological defects}
Let us summarize the topological defects $Top_{GTVW}$ of this model that can be constructed using the techniques described in\ref{Section:4}.  \\
We begin by constructing the lattice $\Gamma^{4,20}$ of RR D-brane charges. Using an orthonormal basis of $\mathbb{R}^{4,20}$, where the basis vectors  $\left\lbrace \vert 1, i \rangle , \vert 2, i \rangle , \vert 3, i \rangle , \vert 4, i \rangle   \right\rbrace_{i=1,2,...,6}$ correspond to the six tetrads described previously, the lattice $\Gamma^{4,20}$  is spanned by the columns of the following matrix:  
\be \arraycolsep=3pt \def\arraystretch{0.8}
\frac{1}{\sqrt{8}}
\left(
\begin{array}{cccccccccccccccccccccccc}
     8 & 4 & 4 & 4 & 4 & 4 & 4 & 2 & 4 & 4 & 4 & 2 & 4 & 2 & 2 & 2 & 4 & 2 & 2 & 2 & 2 & 0 &
   0 & 1 \\
 0 & 4 & 0 & 0 & 0 & 0 & 0 & 2 & 0 & 0 & 0 & 2 & 0 & 2 & 0 & 0 & 0 & 0 & 0 & 2 & 0 & 0 &
   0 & 1 \\
 0 & 0 & 4 & 0 & 0 & 0 & 0 & 2 & 0 & 0 & 0 & 2 & 0 & 0 & 2 & 0 & 0 & 2 & 0 & 0 & 0 & 0 &
   0 & 1 \\
 0 & 0 & 0 & 4 & 0 & 0 & 0 & 2 & 0 & 0 & 0 & 2 & 0 & 0 & 0 & 2 & 0 & 0 & 2 & 0 & 0 & 0 &
   0 & 1 \\
 0 & 0 & 0 & 0 & 4 & 0 & 0 & 2 & 0 & 0 & 0 & 0 & 0 & 2 & 2 & 2 & 0 & 2 & 2 & 2 & 2 & 0 &
   0 & 1 \\
 0 & 0 & 0 & 0 & 0 & 4 & 0 & 2 & 0 & 0 & 0 & 0 & 0 & 2 & 0 & 0 & 0 & 0 & 2 & 0 & 0 & 0 &
   0 & 1 \\
 0 & 0 & 0 & 0 & 0 & 0 & 4 & 2 & 0 & 0 & 0 & 0 & 0 & 0 & 2 & 0 & 0 & 0 & 0 & 2 & 0 & 0 &
   0 & 1 \\
 0 & 0 & 0 & 0 & 0 & 0 & 0 & 2 & 0 & 0 & 0 & 0 & 0 & 0 & 0 & 2 & 0 & 2 & 0 & 0 & 0 & 0 &
   0 & 1 \\
 0 & 0 & 0 & 0 & 0 & 0 & 0 & 0 & 4 & 0 & 0 & 2 & 0 & 2 & 2 & 2 & 0 & 2 & 2 & 2 & 2 & 2 &
   2 & 1 \\
 0 & 0 & 0 & 0 & 0 & 0 & 0 & 0 & 0 & 4 & 0 & 2 & 0 & 2 & 0 & 0 & 0 & 2 & 0 & 0 & 0 & 2 &
   0 & 1 \\
 0 & 0 & 0 & 0 & 0 & 0 & 0 & 0 & 0 & 0 & 4 & 2 & 0 & 0 & 2 & 0 & 0 & 0 & 2 & 0 & 0 & 0 &
   2 & 1 \\
 0 & 0 & 0 & 0 & 0 & 0 & 0 & 0 & 0 & 0 & 0 & 2 & 0 & 0 & 0 & 2 & 0 & 0 & 0 & 2 & 0 & 0 &
   0 & 1 \\
 0 & 0 & 0 & 0 & 0 & 0 & 0 & 0 & 0 & 0 & 0 & 0 & 4 & 2 & 2 & 2 & 0 & 0 & 0 & 0 & 2 & 2 &
   2 & 1 \\
 0 & 0 & 0 & 0 & 0 & 0 & 0 & 0 & 0 & 0 & 0 & 0 & 0 & 2 & 0 & 0 & 0 & 0 & 0 & 0 & 0 & 2 &
   0 & 1 \\
 0 & 0 & 0 & 0 & 0 & 0 & 0 & 0 & 0 & 0 & 0 & 0 & 0 & 0 & 2 & 0 & 0 & 0 & 0 & 0 & 0 & 0 &
   2 & 1 \\
 0 & 0 & 0 & 0 & 0 & 0 & 0 & 0 & 0 & 0 & 0 & 0 & 0 & 0 & 0 & 2 & 0 & 0 & 0 & 0 & 0 & 0 &
   0 & 1 \\
 0 & 0 & 0 & 0 & 0 & 0 & 0 & 0 & 0 & 0 & 0 & 0 & 0 & 0 & 0 & 0 & 4 & 2 & 2 & 2 & 2 & 2 &
   2 & 1 \\
 0 & 0 & 0 & 0 & 0 & 0 & 0 & 0 & 0 & 0 & 0 & 0 & 0 & 0 & 0 & 0 & 0 & 2 & 0 & 0 & 0 & 2 &
   0 & 1 \\
 0 & 0 & 0 & 0 & 0 & 0 & 0 & 0 & 0 & 0 & 0 & 0 & 0 & 0 & 0 & 0 & 0 & 0 & 2 & 0 & 0 & 0 &
   2 & 1 \\
 0 & 0 & 0 & 0 & 0 & 0 & 0 & 0 & 0 & 0 & 0 & 0 & 0 & 0 & 0 & 0 & 0 & 0 & 0 & 2 & 0 & 0 &
   0 & 1 \\
 0 & 0 & 0 & 0 & 0 & 0 & 0 & 0 & 0 & 0 & 0 & 0 & 0 & 0 & 0 & 0 & 0 & 0 & 0 & 0 & 2 & 2 &
   2 & 1 \\
 0 & 0 & 0 & 0 & 0 & 0 & 0 & 0 & 0 & 0 & 0 & 0 & 0 & 0 & 0 & 0 & 0 & 0 & 0 & 0 & 0 & 2 &
   0 & 1 \\
 0 & 0 & 0 & 0 & 0 & 0 & 0 & 0 & 0 & 0 & 0 & 0 & 0 & 0 & 0 & 0 & 0 & 0 & 0 & 0 & 0 & 0 &
   2 & 1 \\
 0 & 0 & 0 & 0 & 0 & 0 & 0 & 0 & 0 & 0 & 0 & 0 & 0 & 0 & 0 & 0 & 0 & 0 & 0 & 0 & 0 & 0 &
   0 & 1 
   \end{array}
   \right).
\label{GTVW:RR_lattice}
\ee
Since the first tetrad contains the spectral flow generators spanning  $\Pi \in \mathbb{R}^{4,20}$ and the first four vectors of the matrix lie in this subspace, we have $\Pi \cap \Gamma^{4,20} \neq 0$. By Claim \ref{th:qdim}, this implies all defects in the model have integral quantum dimension. 

\begin{itemize}
\item \textit{Invertible defects:} The simplest class of defects in $Top_{GTVW}$ are the invertible ones. In particular, for each group element $g \in G_{GTVW}$, there exists a corresponding simple defect $\mathcal{L}_g$. 
\item \textit{Continuous of defects:}  Using the torus orbifold description $\mathcal{T}/ \mathbb{Z}_2$, we obtain a continuous family of non-invertible defects $T_{\theta} = W_{\theta} + W_{- \theta}$, derived from the invertible defect $W_{\theta}$ of the torus, for each $\theta \in \left( \Gamma^{4,4} \otimes \mathbb{R} \right) / \Gamma^{4,4}$, where $\Gamma^{4,4}$ is the Narain lattice of $\mathcal{T}$. These defects have quantum dimension $2$. Identifying the first two tetrads with the untwisted sectors, the defects $T_{\theta}^{(2)}$ act trivially on these sectors, namely by multiplication by the quantum dimension, while annihilating the remaining sectors. Furthermore, since invertible defects in $G_{GTVW}$ exchange the second tetrad with one of the last four while leaving the first tetrad (associated with spectral flow) invariant, fusing these symmetries with the defects $T_{\theta}$ yields additional continuous families of defects with quantum dimension $2$. These new defects act trivially on the spectral flow generators and another tetrad while annihilating the remaining ones. \\
Additionally, since the model admits a $\mathcal{T}/ \mathbb{Z}_4$  orbifold description, we can construct a continuum of topological defects $T^{(4)}_{\theta}$ of quantum dimension $4$, and a continuum of defects $\xi^{(2)}_{\theta}$ of quantum dimension $2$, both of which act trivially on the first tetrad and annihilate the remaining five tetrads. 
\item Since $\mathcal{C}_{GTVW}$ is a RCFT with respect to the chiral and anti-chiral algebras $(\hat{su}(2)_1)^{\oplus 6}$, one can construct topological defects in the bosonic CFT $\mathcal{C}_{GTVW}^{bos}$ that preserve this algebra. These are \textit{Verlinde lines}. \\
Verlinde defects are in one-to-one correspondence with the irreducible representations of the algebra $\mathcal{A} \times \bar{A}$ and their fusion ring matches the fusion algebra generated by the tensor product decomposition of these representations. Each $\hat{su}(2)_1$ term has two irreducible representations, leading to $2^6$ Verlinde lines for each $(\hat{su}(2)_1)^{\oplus 6}$,  which obey a group-like fusion algebra $\mathbb{Z}_2^6$. Consequently, all of these defects are invertible. When lifted to the fermionic $\mathcal{C}_{GTVW}$ SCFT, these defects generate only a subgroup $\left( \mathbb{Z}_2^6 \times \mathbb{Z}_2^6 \right) / \mathbb{Z}_2^5 \cong \mathbb{Z}_2^7$. These lines only preserve the even subalgebra of $\mathcal{N}=4$ but not the supercurrents. The subset of Verlinde lines that do preserve the full superalgebra, and thus correspond to genuine defects in $Top_{GTVW}$, is given by the intersection: $G_{GTVW} \cap \mathbb{Z}_2^7 \cong \mathbb{Z}^4_2$. \\
A natural generalization of this analysis involves considering defects that preserve a smaller (but still rational) subalgebra of $ \mathcal{A} \times \bar{\mathcal{A}}$, potentially leading to a broader class of topological defects.
\item An alternative approach is to consider defects that do not fix the $(\hat{su}(2)_1)^{\oplus 6}$ chiral and anti-chiral algebras,  but instead act on them via a non-trivial automorphism. Such defects implement an action $( \rho_R , \rho_L)$  (not necessarily symmetric) on the algebras, scaled by the quantum dimension. The automorphism group of  $(\hat{su}(2)_1)^{\oplus 6}$ is $SU(2)^6 \rtimes S^6$, where $SU(2)^6$ corresponds to inner automorphisms (generated by current zero modes), while $S^6$ corresponds to outer automorphisms (permutations of the $su(2)_1$ factors). In order for the defect to preserve, beyond the chiral algebra, also the supercurrents, $\rho_{R/L}$  must belong to a finite subgroup of the automorphism group. If the automorphism $(\rho_L, \rho_R)$ lifts to a CFT symmetry, the defect is an invertible Verlinde line. New non-invertible defects arise when the automorphism does not lift to a CFT symmetry. These correspond to non-trivial double cosets in
{\small{
\begin{equation}
 \left( (SU(2)^6 \times SU(2)^6) \rtimes S_6^{diag} \right) \backslash  \left( (SU(2)^6 \times SU(2)^6) \rtimes (S_6 \times S_6) \right) / \left( (SU(2)^6 \times SU(2)^6) \rtimes S_6^{diag} \right) .
\end{equation}}}
Each coset can be represented by an element $(\rho_L, \rho_L)$ with $\rho_R = 1$. \\
The requirement to preserve the super-algebra $\mathcal{N}=(4,4)$ implies the $S_6$ permutation to be contained in the subgroup $A_5 \subset S_6$, which contains only three non-trivial cycle shapes, corresponding to the partitions $1+1+1+3$, $1+1+2+2$ or $1+5$. For each of these partitions one can consider a representative automorphism action for $\rho_L$ and construct a non-invertible duality defect implementing this action. Up to conjugation with elements in $G_{GTVW}$, the three classes contain the following defects: $\mathcal{N}_{ijk}$ for all $2 \leq i < j < k \leq 6$, with $\mathcal{N}^2_{ ijk} = \mathcal{I}+\mathcal{L}_{t_it_j} +\mathcal{L}_{t_jt_k} +\mathcal{L}_{t_it_k} $; $\mathcal{N}_{ij,kl}$ for all pairwise distinct $i, j, k, l \in {2,...,6}$, with $\mathcal{N}^2_{ ij,kl} = \mathcal{I} + \mathcal{L}_{t_it_j} + \mathcal{L}_{t_kt_l} + \mathcal{L}_{t_it_jt_kt_l}$ ; and $\mathcal{N}_{23456}$ where $\mathcal{N}^2_{ 23456}$ equals to the superposition of the 16 invertible defects in $G_{GTVW}  \cap (\mathbb{Z}^6_2  \times \mathbb{Z}^6_2)/ \mathbb{Z}^5_2 \cong \mathbb{Z}_2^4$. For each of these  defects  one can construct an explicit endomorphism   \eqref{homtoB} acting on the lattice \eqref{GTVW:RR_lattice}. For example, for the defect $\mathcal{N}_{256}$ we have the following explicit action:
\be\arraycolsep=3pt \def\arraystretch{0.8}\mathsf{N}_{256}=
\left(
\begin{array}{cccccccccccccccccccccccc}
 2 & 0 & 0 & 0 & 0 & 0 & 0 & 0 & 0 & 0 & 0 & 0 & 0 & 0 & 0 & 0 & 0 & 0 & 0 & 0 & 0 & 0 & 0 & 0 \\
 0 & 2 & 0 & 0 & 0 & 0 & 0 & 0 & 0 & 0 & 0 & 0 & 0 & 0 & 0 & 0 & 0 & 0 & 0 & 0 & 0 & 0 & 0 & 0 \\
 0 & 0 & 2 & 0 & 0 & 0 & 0 & 0 & 0 & 0 & 0 & 0 & 0 & 0 & 0 & 0 & 0 & 0 & 0 & 0 & 0 & 0 & 0 & 0 \\
 0 & 0 & 0 & 2 & 0 & 0 & 0 & 0 & 0 & 0 & 0 & 0 & 0 & 0 & 0 & 0 & 0 & 0 & 0 & 0 & 0 & 0 & 0 & 0 \\
 0 & 0 & 0 & 0 & 0 & 0 & 0 & 0 & 0 & 0 & 0 & 0 & 0 & 0 & 0 & 0 & 0 & 0 & 0 & 0 & 0 & 0 & 0 & 0 \\
 0 & 0 & 0 & 0 & 0 & 0 & 0 & 0 & 0 & 0 & 0 & 0 & 0 & 0 & 0 & 0 & 0 & 0 & 0 & 0 & 0 & 0 & 0 & 0 \\
 0 & 0 & 0 & 0 & 0 & 0 & 0 & 0 & 0 & 0 & 0 & 0 & 0 & 0 & 0 & 0 & 0 & 0 & 0 & 0 & 0 & 0 & 0 & 0 \\
 0 & 0 & 0 & 0 & 0 & 0 & 0 & 0 & 0 & 0 & 0 & 0 & 0 & 0 & 0 & 0 & 0 & 0 & 0 & 0 & 0 & 0 & 0 & 0 \\
 0 & 0 & 0 & 0 & 0 & 0 & 0 & 0 & 0 & 0 & 0 & 0 & 1 & -1 & -1 & -1 & 0 & 0 & 0 & 0 & 0 & 0 & 0 & 0
   \\
 0 & 0 & 0 & 0 & 0 & 0 & 0 & 0 & 0 & 0 & 0 & 0 & 1 & 1 & -1 & 1 & 0 & 0 & 0 & 0 & 0 & 0 & 0 & 0 \\
 0 & 0 & 0 & 0 & 0 & 0 & 0 & 0 & 0 & 0 & 0 & 0 & 1 & 1 & 1 & -1 & 0 & 0 & 0 & 0 & 0 & 0 & 0 & 0 \\
 0 & 0 & 0 & 0 & 0 & 0 & 0 & 0 & 0 & 0 & 0 & 0 & 1 & -1 & 1 & 1 & 0 & 0 & 0 & 0 & 0 & 0 & 0 & 0 \\
 0 & 0 & 0 & 0 & 0 & 0 & 0 & 0 & 1 & 1 & 1 & 1 & 0 & 0 & 0 & 0 & 0 & 0 & 0 & 0 & 0 & 0 & 0 & 0 \\
 0 & 0 & 0 & 0 & 0 & 0 & 0 & 0 & -1 & 1 & 1 & -1 & 0 & 0 & 0 & 0 & 0 & 0 & 0 & 0 & 0 & 0 & 0 & 0 \\
 0 & 0 & 0 & 0 & 0 & 0 & 0 & 0 & -1 & -1 & 1 & 1 & 0 & 0 & 0 & 0 & 0 & 0 & 0 & 0 & 0 & 0 & 0 & 0 \\
 0 & 0 & 0 & 0 & 0 & 0 & 0 & 0 & -1 & 1 & -1 & 1 & 0 & 0 & 0 & 0 & 0 & 0 & 0 & 0 & 0 & 0 & 0 & 0 \\
 0 & 0 & 0 & 0 & 0 & 0 & 0 & 0 & 0 & 0 & 0 & 0 & 0 & 0 & 0 & 0 & 0 & 0 & 0 & 0 & 0 & 0 & 0 & 0 \\
 0 & 0 & 0 & 0 & 0 & 0 & 0 & 0 & 0 & 0 & 0 & 0 & 0 & 0 & 0 & 0 & 0 & 0 & 0 & 0 & 0 & 0 & 0 & 0 \\
 0 & 0 & 0 & 0 & 0 & 0 & 0 & 0 & 0 & 0 & 0 & 0 & 0 & 0 & 0 & 0 & 0 & 0 & 0 & 0 & 0 & 0 & 0 & 0 \\
 0 & 0 & 0 & 0 & 0 & 0 & 0 & 0 & 0 & 0 & 0 & 0 & 0 & 0 & 0 & 0 & 0 & 0 & 0 & 0 & 0 & 0 & 0 & 0 \\
 0 & 0 & 0 & 0 & 0 & 0 & 0 & 0 & 0 & 0 & 0 & 0 & 0 & 0 & 0 & 0 & 0 & 0 & 0 & 0 & 0 & 0 & 0 & 0 \\
 0 & 0 & 0 & 0 & 0 & 0 & 0 & 0 & 0 & 0 & 0 & 0 & 0 & 0 & 0 & 0 & 0 & 0 & 0 & 0 & 0 & 0 & 0 & 0 \\
 0 & 0 & 0 & 0 & 0 & 0 & 0 & 0 & 0 & 0 & 0 & 0 & 0 & 0 & 0 & 0 & 0 & 0 & 0 & 0 & 0 & 0 & 0 & 0 \\
 0 & 0 & 0 & 0 & 0 & 0 & 0 & 0 & 0 & 0 & 0 & 0 & 0 & 0 & 0 & 0 & 0 & 0 & 0 & 0 & 0 & 0 & 0 & 0 \\
\end{array}
\right).
\ee
When a group $H$ acts non-trivially on the chiral algebra, self-orbifold consistency requires holomorphic currents in twisted sectors. Analysis of  $G_{GTVW}$ reveals three relevant conjugacy classes: (1) $\mathbb{Z}_2$-type symmetries  yielding to torus models distinct from $\mathcal{C}_{GTVW}$; (2) order-4 symmetries 
 interpretable as $\mathcal{C}_{GTVW}/ \mathbb{Z}_4$ quantum symmetries, again producing torus models; (3) their holomorphic/anti-holomorphic swaps. While $\mathbb{Z}_2$ or $\mathbb{Z}_4$ orbifolds break self-duality, larger groups (e.g., $\mathbb{Z}_2 \times \mathbb{Z}_8$) give rise to duality defects  of order 4 or $8$ that preserve it, though a full classification remains open.\\
The final outcome is that, in addition to invertible defects, one also finds non-invertible ones—such as the duality defects discussed above—which arise from self-orbifold constructions of the form $\mathcal{C}_{GTVW} \cong \mathcal{C}_{GTVW} / H$, for certain subgroups $H$ of $G_{GTVW} \cap \left( \mathbb{Z}_2^6 \times \mathbb{Z}_2^6 \right) / \mathbb{Z}_2^5$. These duality defects belong to the set $T^{(2)}_{\theta}$ obtained via a torus orbifold construction of the model at special values of the parameter $\theta$. 
\end{itemize}

\section{Conclusions and outlook}
\label{sec:conclusions}
In this note we have discussed topological defects in $K3$ non-linear sigma models along the lines of \cite{Angius:2024evd}. While for general non-rational theories, characterizing these objects is problematic due to the absence of strong constraints, in the case of $K3$ models we have shown that a quite straightforward generalization of the lattice-theoretic approach used to classify ordinary symmetries  can lead to interesting general results for the categories $\cTop_{\calC}$ of topological defects preserving the $\CN=(4,4)$ superconformal algebra and the spectral flow in a supersymmetric non-linear sigma model on K3 $\calC$.  For example, one can show that the set of $K3$ models where $\cTop$ is not trivial (i.e., where there are simple defects distinct from the identity) has zero measure in the moduli space. Furthermore, one can restrict the possible quantum dimensions of the defects in $\cTop$ and give a sufficient condition on the points in the moduli space where all quantum dimensions are integral. The method is also valuable in studying specific models such as the model $\mathcal{C}_{GTVW}$ with maximal symmetry $\ZZ_2^8:M_{20}$. We point out the classification is not complete since the map between $\cTop_\Pi$ and the corresponding space of the linear endomorphisms of the D-brane charge lattice $\Gamma^{4,20}$ is neither surjective nor injective. Probably, further conditions must be imposed on TDLs to fully characterize them. A full proof of the conjecture \ref{th:onlytori} that any $K3$ model $\calC$ exhibiting a continuum of topological defects   preserving the $\CN=(4,4)$ superconformal algebra and the spectral flow is still missing because the argument is based on an assumption about the consistency of a generalized orbifold, which in general can fail. Therefore it would be interesting to study examples of $K3$ models that are not generalized orbifolds of torus models in order to find possible counterexamples.

Recently, generalized symmetries in $V^{f\natural}$ have been studied with a similar lattice-theoretic approach \cite{Volpato:2024goy,Angius:2025xxx,Angius:2025zlm}.
We point out the importance of a thorough classification of topological defects in $K3$ models in light of the recent conjectured generalization of  the mysterious correspondence observed in \cite{Duncan:2015xoa} between symmetries of K3 sigma models and automorphisms of the $\CN=1$ supersymmetric vertex operator algebra $V^{f\natural}$ with central charge $12$\footnote{The correspondence was initially formulated in terms of the SVOA $V^{s \natural}$, which is isomorphic to $V^{f \natural}$, but it carries a different action of the Conway group $Co_0$}. 

{\bf Acknowledgements.} We are thankful to Sarah M. Harrison and Roberto Volpato for collaboration and discussions on related research. S.G. would like to thank the organizers of the AMS-UMI  joint conference and in particular the co-organizers Darlayne Addabbo and Lisa Carbone of the special session on ``New Developments in infinite dimensional Lie algebras, vertex operator algebras and the Monster", held in Palermo on July 25-26 2024, for whose proceedings this contribution has been prepared. He is also grateful to all the participants of the session for relevant discussions. R.A. is supported by the ERC Starting Grant QGuide-101042568- StG 2021. This research of S.G has been supported by a BIRD-2021 project (PRD-2021) and by the  PRIN Project n. 2022ABPBEY, ``Understanding quantum field theory through its deformations''. 
 
\printbibliography

\end{document}